\documentclass[fleqn,usenatbib]{mnras}

\usepackage{newtxtext,newtxmath}
\usepackage{amsmath, color,graphicx,tablefootnote,subfigure}
\usepackage{graphics,subfigure,hyperref,float}
\usepackage[rightcaption]{sidecap}
\usepackage{array}
\usepackage{pdflscape}
\usepackage{float}
\usepackage[T1]{fontenc}

\DeclareRobustCommand{\VAN}[3]{#2}
\let\VANthebibliography\thebibliography
\def\thebibliography{\DeclareRobustCommand{\VAN}[3]{##3}\VANthebibliography}

\usepackage{graphicx}	% Including figure files
\usepackage{amsmath}	% Advanced maths commands
\usepackage{multirow}
\newcommand{\pcm}{\,cm$^{-2}$}	% per cm-squared
\newcommand{\fesclyc}{$f_{\rm esc}^{\rm LyC}$}	% per cm-squared
\newcommand{\Lya}{Ly$\alpha$}	% per cm-squared

\title[Minor impact of two $z\sim7.5$ galaxies to the EoR]{A negligible contribution of two luminous $z\sim7.5$ galaxies to the ionizing photon budget of reionization}

\author[S. Gazagnes et al.]{
S. Gazagnes$^{1}$,
J. Chisholm$^{1}$,
R. Endsley$^{1}$,
D. A. Berg$^{1}$,
F. Leclercq$^{1}$,
N. Jurlin$^{1}$,
A. Saldana-Lopez$^{2}$,\newauthor
S. L. Finkelstein$^{1}$,
S. R. Flury$^{3}$,
N. G. Guseva$^{4}$,
A. Henry$^{5}$ ,
Y. I. Izotov$^{4}$,
I. Jung$^{5}$,
J. Matthee$^{6}$,
D. Schaerer$^{7,8}$
\\
$^{1}$Department of Astronomy, The University of Texas at Austin, 2515 Speedway, Stop C1400, Austin, TX 78712-1205, USA\\
$^{2}$Department of Astronomy, Oskar Klein Centre, Stockholm University, 106 91 Stockholm, Sweden\\
$^{3}$Department of Astronomy, University of Massachusetts Amherst, Amherst, MA 01002, United States\\
$^{4}$Bogolyubov Institute for Theoretical Physics, National Academy of
Sciences of Ukraine, 14-b Metrolohichna str., Kyiv, 03143, Ukraine\\
$^{5}$Space Telescope Science Institute, 3700 San Martin Drive Baltimore, MD 21218, United States\\
$^{6}$Institute of Science and Technology Austria (ISTA), Am Campus 1, A-3400 Klosterneuburg, Austria\\
$^{7}$Observatoire de Gen\`eve, Universit\'e de Gen\`eve, Chemin
Pegasi 51, 1290 Versoix, Switzerland\\
$^{8}$CNRS, IRAP, 14 Avenue E. Belin, 31400 Toulouse, France\\
}

\date{Accepted XXX. Received YYY; in original form ZZZ}

\pubyear{2024}

\begin{document}
\label{firstpage}
\pagerange{\pageref{firstpage}--\pageref{lastpage}}
\maketitle

\begin{abstract}
We present indirect constraints on the absolute escape fraction of ionizing photons (\fesclyc) of the system GN 42912 which comprises two luminous ($M_{\rm UV}$ magnitudes of $-20.89$ and $-20.37$) galaxies at $z\sim7.5$, GN~42912-NE and GN~42912-SW, to determine their contribution to the ionizing photon budget of the Epoch of Reionization (EoR). The high-resolution {\sl James Webb Space Telescope} ({\sl JWST}) NIRSpec and NIRCam observations reveal they are separated by only $\sim0\farcs1$ (0.5 kpc) on the sky and have a 358 km s$^{-1}$ velocity separation. GN~42912-NE and GN~42912-SW are relatively massive for this redshift (log($M_\ast/M_\odot$) $\sim$ 8.4 and 8.9, respectively), with gas-phase metallicities of 18 per cent and 23 per cent solar, O$_{32}$ ratios of 5.3 and $>5.8$, and $\beta$ slopes of $-1.92$ and $-1.51$, respectively. We use the \ion{Mg}{ii}~$\lambda\lambda$2796,2803 doublet to constrain \fesclyc. \ion{Mg}{ii} has an ionization potential close to that of neutral hydrogen and, in the optically thin regime, can be used as an indirect tracer of the LyC leakage. We establish realistic conservative upper limits on \fesclyc\ of 8.5 per cent for GN~42912-NE and 14 per cent for GN~42912-SW. These estimates align with \fesclyc\ trends observed with $\beta$, O$_{32}$, and the H$\beta$ equivalent width at $z<4$. The small inferred ionized region sizes ($<0.3$ pMpc) around both galaxies indicate they have not ionized a significant fraction of the surrounding neutral gas. While these $z>7$ \fesclyc\ constraints do not decisively determine a specific reionization model, they support a minor contribution from these two relatively luminous galaxies to the EoR.
\end{abstract}

% Don't make up new ones.We constrain \fesclyc\ in both galaxies adopting multiple scenarios to accommodate uncertainties regarding dust attenuation.

%%%%%%%%%%%%%%%%%%%%%%%%%%%%%%%%%%%%%%%%%%%%%%%%%%

%%%%%%%%%%%%%%%%% BODY OF PAPER %%%%%%%%%%%%%%%%%%

\section{Introduction}
\label{sec:int}

The Epoch of Reionization (EoR) is a significant phase transition in the Universe \citep{barkana2001, furlanetto2006}. After recombination, the hydrogen gas within the inter-galactic medium (IGM) existed in a neutral state. The formation of the first stars, galaxies, and active galactic nuclei (AGN) led to the production and propagation of a sufficient quantity of ionizing photons (photons with $\lambda<$ 912\AA, also known as Lyman Continuum or LyC photons) to reionize the neutral hydrogen in the circum-galactic medium (CGM) and IGM \citep{dayal2018, dayal2020}. This intricate process remains largely enigmatic, with significant debate over which sources (e.g. star-forming galaxies, AGN, intermediate-mass black holes) contributed most to the total ionizing budget needed to explain reionization \citep{robertson, madau2015, rosdahl2018, finkelstein2019}. 

To gauge the contribution of the primary sources of ionizing photons, we must determine their ionizing emissivity, i.e., the rate at which they emit ionizing photons per unit time and volume \citep{madau1999, Miralda2000_reio}. The combined emissivity of all sources must be sufficient to ionize the hydrogen atoms in the IGM while accounting for potential recombination events. The emissivity of a source of ionizing photons, $\dot{n}_{\rm ion}$, is derived as

\begin{equation}
    \dot{n}_{\rm ion} = \rho_{\rm UV}\times \xi_{\rm ion} \times \text{\fesclyc}
\end{equation}

\noindent where $\rho_{\rm UV}$ is the relative density of the emitting sources across various UV brightness bins to some assumed limiting luminosity, $\xi_{\rm ion}$ is the intrinsic production of ionizing photons per unit non-ionizing UV luminosity, and \fesclyc\ is the fraction of ionizing photons that successfully escapes the interstellar medium (ISM)  and CGM of galaxies (\fesclyc) and contributes to the IGM-hydrogen reionization.

Prior to the {\sl James Webb Space Telescope} ({\sl JWST}) era, studies extrapolated post-reionization observations to formulate the relevant reionization models \citep[e.g.][]{robertson,robertson2015, madau2015,finkelstein2019, naidu2020, matthee2022_lyc, chisholm2022_beta}. These models aimed to satisfy specific constraints, such as the opacity of the Cosmic Microwave Background ($\tau_{\rm CMB}$, \citealt{planck2016}), and observations of damped Lyman-$\alpha$ (\Lya) wings in quasars \citep[e.g.][]{greig2019_lyadamping} which provides insights into neutral gas fractions at $z>6$.

In the pre-{\sl JWST} era, several studies favored star-forming galaxies (SFGs) as the dominant sources of cosmic reionization \citep[e.g.][]{ouchi, robertson, finkelstein2012_candels, naidu2020}. SFG-based reionization models are typically separated into faint galaxy and bright galaxy models \citep[e.g.][]{greig2015}. In the latter scenario, rare luminous galaxies, located in the highest-density regions of the universe, generate a substantial number of ionizing photons and dominate the ionizing budget of the EoR \citep{Marques-Chaves2022_bluenugget, naidu2020}. In contrast, the faint-galaxy scenario builds upon galaxies found in smaller haloes, more numerous but producing fewer ionizing photons individually. While both types of models match the current constraints on $\tau_{\rm CMB}$ and the neutral gas fraction at $z>6$, they have significantly distinct implications for the timeline, morphology, and structure of cosmic reionization \citep{robertson, finkelstein2019, kanna2022_thesan, gazagnes2021_inferring}.

The advent of the {\sl JWST} has offered unique insights into the high-redshift ($z>6$) universe. Interestingly, while SFGs-dominated models were favored, {\sl JWST} observations unveiled an unexpected abundance of faint AGN at $z>5$ \citep[e.g.][]{maioline2023_bh, maiolino2024_natbh, matthee2024, Larson2023_agnceers, kocevski2023_agn, kocevski2024_agn2, furtak2024_agn}. While these AGN are predominantly dust-reddened \citep{dayal2024_agn_reio, casey2024_dustagn}, their contribution to reionization is still debated \citep{madau2024_agn}. Additionally, {\sl JWST} observations introduced new complexities to the overall picture of SFG-based reionization models.  {\sl JWST} unveiled super-early luminous galaxies at redshifts greater than 9 \citep{donnan2023_massive, finkelstein2024_massive, harikane2023_earlyjwst}, as well as galaxies at redshifts greater than 6 exhibiting a relative high ionizing photon production (${\rm log}\ \xi_{\rm ion}$ $\geq$ 25.5) \citep{Atek2024_epsion, simmonds2024, endsley2023, prieto-lyon2023, hsiao2024} and star formation activity \citep{finkelstein2023_sfrreio, harikane2023_earlyjwst, eisenstein2023_jades}. This observed increase in $\xi_{\rm ion}$ and $\rho_{\rm UV}$ at high-redshift results suggests there may be
too many ionizing photons escaping SFGs at $z>6$, leading to a reionization timeline incompatible with the current constraints on $\tau_{\rm CMB}$ and the neutral gas fraction at $z>6$ \citep{munoz2024}. 

% Ultimately, these models established straightforward predictions for the reionization process, which could be tested with current and upcoming instruments probing the high-redshift universe.although several studies continue to support the dominant role of SFGs \citep[e.g.][]{Atek2024_epsion},  have introduced new complexities to the overall picture.

% To explain reionization, several pre-{\sl JWST} studies built upon the hypothesis that star-forming galaxies (SFGs) played a more significant role than AGN in cosmic reionization \citep[e.g.][]{ouchi, robertson, finkelstein2012_candels, naidu2020}.

% Indeed, an increase in $\xi_{\rm ion}$ and $\rho_{\rm UV}$ at redshifts greater than 6 results in an overestimation of the ionizing emissivity of the primary ionizing sources in both faint and bright galaxy models,  .
%These relations are empirically derived from $z<4$ samples of LyC-leaking galaxies , yet there is no \fesclyc-to-$M_{\rm UV}$ relation in the post-reionization era that adequately describes the diversity of LyC leakers observed \citep[e.g.][]{Jung2024_lowescape, Marques-Chaves2022_bluenugget}. 

In light of the new insights provided by {\sl JWST}, determining the distribution of \fesclyc\ across SFGs has become a critical need. Indeed, both faint and bright galaxy models rely on different \fesclyc-to-$M_{\rm UV}$ relations \citep[e.g.][]{chisholm2022_beta, matthee2022_lyc}. Furthermore, the observation of a lower average \fesclyc\ than assumed by pre-{\sl JWST} reionization models (3\% instead of 5-10\%) could reconcile these models with current constraints on $\xi_{\rm ion}$ and $\rho_{\rm UV}$ \citep{munoz2024}. Thus, determining the \fesclyc\ of SFGs is pivotal for understanding the process and timeline of reionization \citep{sharma2016_timeline, naidu2020, finkelstein2019}.

% premay be the balancing parameter in the context of SFGs of larger $\xi_{\rm ion}$ and $\rho_{\rm UV}$ during reionization. of SFGs during reionization may help understand  reconcile pre-{\sl JWST} reionization models with the current constraints on cosmic reionization. Indeed, Finding a lower average \fesclyc\ (

Currently, few studies have set tight constraints on the \fesclyc\ of reionization-era objects \citep[e.g.][]{mascia2024_feschighz, jaskot2024_lycmultiII}. This is because robustly constraining \fesclyc\ at high redshift is virtually impossible. It necessitates direct observations below 912\AA, which are unfeasible at redshifts greater than 4 due to IGM absorption \citep{worseck2014, vanzella2015}. Hence, our best approach for constraining \fesclyc\ relies on indirect diagnostics, established from the $z<4$ Universe, where direct constraints can be made and compared to other UV and optical properties observable with {\sl JWST}. Fortunately, the past decade has seen an exponential growth in the discovery of LyC leaking galaxies at $z<6$ \citep{leitet2013, borthakur2014, leitherer2016, izotov2016a, izotov2018a, izotov2016b, izotov2018b, izotov21_lowmassLYC,vanzella2015, debarros2016, shapley2016, bian2017, steidel2018, fletcher2019, rivera2019_lycsunburst, pahl2021_lyc}. In particular, the Low-$z$ Lyman Continuum Survey (LzLCS, \citealt{flury2022_lzlcs}) has substantially augmented the number of LyC detections, adding 35 new LyC galaxies at $z<0.5$ and enhancing the diversity and completeness of the sample. This endeavor has advanced our understanding of ionizing photon escape and laid the groundwork for establishing LyC diagnostics applicable to high-$z$ studies \citep{saldana2022, flury2022_lycdiag, leclercq2024_lycMgII, amorin2024, chisholm2022_beta, bait2024_radiolzlcs, wang2021_SII}.

Both theoretical and observational studies of LyC leaking galaxies emphasized that the properties of neutral gas, particularly its density, and geometry, along with dust extinction, are key regulators of the ionizing leakage \citep[e.g.][]{gazagnes2018, gazagnes2020, kimm2017, chisholm2020}. Consequently, spectral features and properties that trace neutral gas and dust (e.g. \Lya, $\beta$ slopes, low ionization states of metal lines) are generally robust indicators of LyC escape. Importantly, the best diagnostics should combine insights into both neutral gas and dust properties \citep{chisholm2022_beta, gazagnes2024_vandels}. 

% However, in practice, lines of sight cleared of neutral gas tend to also be dust-free \citep{gazagnes2024_vandels} such that diagnostics based on either neutral gas or dust properties are still reliable in identifying LyC-leaking candidates.

The \ion{Mg}{ii}$\ \lambda\lambda$2796,2803 doublet is one of the most promising diagnostics for constraining \fesclyc\ at high-$z$. Since the IGM is predominantly neutral at $z>6$, direct neutral gas diagnostics like \Lya\ are heavily impacted by the neutral gas and can only trace the most luminous leakers residing in large ionized bubbles that favor \Lya\ transmission \citep{mason_lya}. \ion{Mg}{ii} presents a powerful alternative; its ionization potential is closely aligned with that of \ion{H}{i} (15 eV versus 13.6 eV), suggesting that \ion{Mg}{ii} can serve as a tracer for neutral gas density and thereby indirectly infer LyC escape \citep{henry2018, chisholm2020, chang2024_mgii}. Observational studies have highlighted a remarkable agreement between LyC escape fraction derived using \ion{Mg}{ii}-based approaches and directly-constrained \fesclyc\ at low-$z$ \citep{chisholm2020,leclercq2024_lycMgII, xu2022_lya_mgII, xu2023_lycmgII}.

% \ion{Mg}{ii} still suffers some caveats, it is metallicity and ionization dependant, and cosmological simulations underscored the complexity of the relationship between LyC leakage and \ion{Mg}{ii} emission due to resonant scattering effects that can alter \ion{Mg}{ii} line properties \citep{katz2022_mgii}. Nevertheless,  

Building upon these results, a Cycle 1 program (ID: 1871, PI: Chisholm) was granted $\sim22.2$ hours of observations to capture the \ion{Mg}{ii} emission of 20 reionization-era galaxies in the GOODS-North field and establish the very first indirect \fesclyc\ constraints of high-redshift galaxies.  In this paper, we analyze the {\sl JWST} Near Infrared Spectrograph \citep[NIRSpec,][]{boker2023_nirspec} high-spectral resolution observations of GN~42912, a bright \Lya\ system at $z$ = 7.5 \citep{finkelstein2013_lyadet, hutchison2019_ciii, jung2020_lyatexas}. Our objective is to establish the first constraints on the absolute \fesclyc\ for this reionization-era system, integrating insights on both neutral gas and dust content. We will evaluate the significance of our findings within the framework of pre-{\sl JWST} reionization models from \citet{chisholm2022_beta} and \citet{matthee2022_lyc} and the current ionizing photon budget crisis \citep{munoz2024}.

This paper is organized as follows: Section~\ref{sec:obs} introduces the observations and reduction strategy of the {\sl JWST} data. In  Section~\ref{sec:meas}, we analyze the Spectral Energy Distribution (SED) properties and \ion{Mg}{ii}, [\ion{O}{iii}], [\ion{O}{ii}], and Balmer emission lines of GN 42912. We present the \ion{Mg}{ii}-based constraints on the \fesclyc\ in  Section~\ref{sec:mgII}. In  Section~\ref{sec:robustness}, we discuss the reliability of these estimates. We compare these constraints to low-$z$ LyC leakers, trends, and alternative \fesclyc\ diagnostics in Section~\ref{sec:lyclit}. Finally, we consider these $z\sim7.5$ \fesclyc\ constraints in the context of reionization models in Section~\ref{sec:reio}. We conclude in  Section~\ref{sec:conc}.

Throughout this paper, we use a cosmology with H$_0$=67.4 km s$^{-1}$ Mpc$^{-1}$ and $\Omega_{\rm M}$=0.315 \citep{planck2016} and the solar metallicity is defined as 12+log(O/H) of 8.69 \citep{asplund2021}. All magnitudes are reported as absolute AB magnitudes and the restframe wavelengths of emission lines quoted are given in Angstroms (\AA) in the vacuum frame using the National Institute of Standards and Technology database \citep[NIST, ][]{NISTASD2022}. The uncertainties on all measurements and equations include the propagation of all the uncertainties involved in each, using the \textsc{python} package \textsc{uncertainties} \citep{uncertainties}.

\section{Observations}
\label{sec:obs}

Here we describe the {\sl JWST} observations of GN~42912. Section~\ref{sec:project} briefly presents the {\sl JWST} Cycle 1 program. Section~\ref{sec:nir} and Section~\ref{sec:nirspec} describe the NIRcam and NIRSpec data reduction.

\subsection{{\sl JWST} Project ID 1871}
\label{sec:project}

This paper focuses on GN~42912, a $z\sim7.5$ system observed in Cycle 1 of {\sl JWST} as part of the {\sl JWST} Project ID 1871 (PI: Chisholm). \citet{chisholm24_nev} describes the details of the program and its data reduction, and we summarize the important steps here. Project ID 1871 selected 20 high-redshift star-forming galaxies within the GOODS-North (GN) field with photometric redshifts from \citet{finkelstein2015_luv}, and some with \Lya-based redshift estimation from \citet{jung2020_lyatexas}. The goal was to capture the velocity profiles of the \ion{Mg}{ii}~$\lambda\lambda$2796,2803 emission lines and deduce the neutral gas column density (and associated LyC escape fraction, \citealt{henry2018, chisholm2020}). The data were acquired from the {\sl JWST} \citep{jwst_telescope_2023, rigby2023_jwst} on February 10, 2023, utilizing the NIRSpec micro-shutter assembly (MSA) G235H/F170LP and G395H/F290LP grating capabilities. The MSA observations were centered on a prominent, bright Ly$\alpha$ emitter at $z = 7.5$ in the GOODS-North field \citep[GN 42912,][]{finkelstein2013_lyadet, hutchison2019_ciii, jung2020_lyatexas}.

Of the 20 selected targets, 9 have the \ion{Mg}{ii} 2796\AA\ and 2803\AA\ emission lines falling inside the NIRSpec gratings coverage. Among these 9 objects, GN~42912 stands out with the most robust detection of the \ion{Mg}{ii} doublet (exceeding $3\sigma$), making it the primary focus of this paper. The other galaxies with \ion{Mg}{ii}  coverage do not show significant \ion{Mg}{ii} detections. The next sections provide a comprehensive overview of the data reduction.

\subsection{NIRcam Data Reduction}
\label{sec:nir}

The NIRCam imaging is from the First Reionization Epoch Spectroscopically Complete Observations  \citep[FRESCO,][]{Oesch2023_fresco} and span the medium-band filters (F182M and F210M) in the short-wavelength channel and the F444W filter in the long-wavelength channel. The total exposure times are 4456, 3522, and 934 seconds for the F182M, F210M, and F444W filters, respectively.

% The FRESCO imaging aimed for a $5\sigma$ detection of a 28.2 mag source, with total exposure times of  These NIRCam images served to validate the NIRSpec data reduction and provided insights into the broad-band properties of GN 42912.

The FRESCO NIRCam images were processed following the methodology from \citet{endsley2023_nircam}, using the {\sl JWST} Science Calibration Pipeline (v1.11.3). As detailed in \citet{chisholm24_nev}, we removed snowball and wisp artifacts, integrating sky flats and wisp templates derived from publicly available data, and took the photometric zero points from \citet{boyer2022_fluxcal} within \textsc{jwst\_1106.pmap}. The $1/f$ noise and 2D background subtraction in the \textsc{*\_cal.fits} files was done on an amplifier-by-amplifier basis, utilizing the \textsc{sep} package \citep{barbary2016_sep}.

We used the {\sl Hubble Space Telescope} ({\sl HST}) Wide Field Camera~3 (WFC3)/F160W reductions of the GN field from the Complete Hubble Archive for Galaxy Evolution (CHArGE, \citealt{kokorev2022_charge}) to align the \textsc{*\_cal.fits} files to the {\sl Gaia} astrometric frame. During the pipeline's final stage, we resampled all NIRCam mosaics to a uniform World Coordinate System with a 30 mas pixel$^{-1}$ scale and then adjusted the images to match the point-spread function (PSF) of the F444W filter using PSFs defined within the FRESCO mosaics.

\begin{figure}
    \centering
    \includegraphics[width = \hsize]{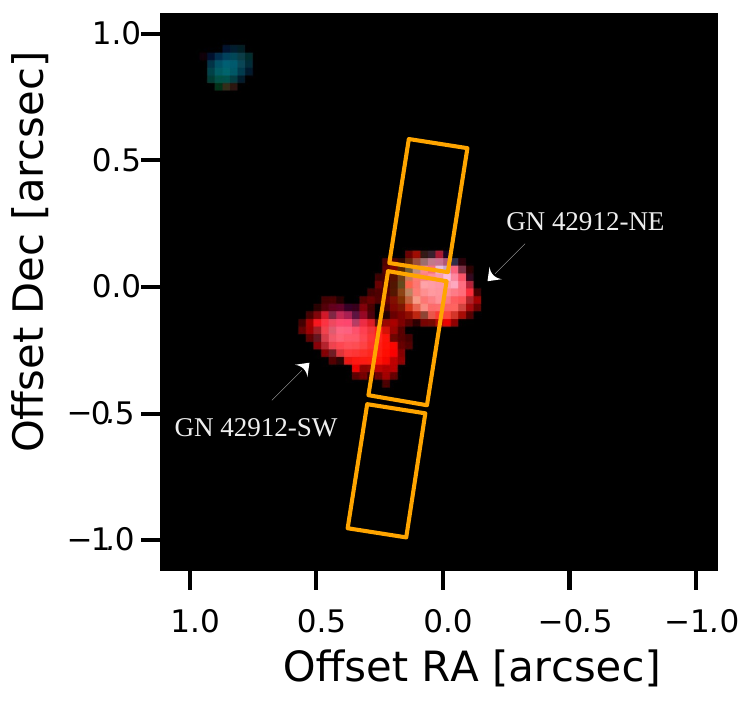}
    \caption{Three-color NIRCam image of GN~42912, with north oriented upwards and east towards the right. The image composite uses the F182M filter for blue, F210M for green, and F444W for red. To compensate for the dominance of the F444W filter in the composite image,  we increased the F182M and F210M by 30 per cent to enhance their appearance. The locations of the three NIRSpec shutters are overlaid in orange. The spatial resolution of the {\sl JWST} filters reveals that GN~42912 comprises of two companion galaxies, not entirely covered by the rectangular NIRSpec shutters. Further details on GN~42912 are provided in Section~\ref{sec:overall}.}
    \label{fig:f444w}
\end{figure}

Figure~\ref{fig:f444w} presents a composite image (F180M, F210M, F444W) of GN~42912. The resolution of the NIRCAM filters can resolve the presence of two components separated by $\sim$ 0\farcs1 on the sky ($\sim 0.5$ kpc). We further detail the properties of both components in Section~\ref{sec:overall}.  In the rest of this paper, we refer to these two components as GN~42912-NE and GN~42912-SW to denote the one in the northeast and the southwest, respectively.

\subsection{NIRSpec Data Reduction}
\label{sec:nirspec}

The NIRSpec observations were divided between the G235H and G395H grating configurations. The G235H observations, focusing on the restframe 2000–3600~\AA\ wavelength range for sources at redshift 7.5, cover the \ion{Mg}{ii}~$\lambda\lambda$2796,2803 emission lines. Given its importance for the program's scientific goals, the G235H grating was allocated a longer exposure time: 53,044 s (approximately 14.7 hours) across 36 exposures. In contrast, the G395H grating, covering the restframe 3400–6000 \AA\ wavelength range for sources at redshift 7.5, contains optical emission lines brighter than \ion{Mg}{ii}, required significantly less exposure time, 9716 seconds (around 2.7 hours) over 6 integrations. Both configurations operated under the NRSIRS2 readout mode, and the standard three-shutter nod pattern was employed to capture sufficient background and enable a robust background subtraction. 

Concerning the NIRSpec reduction pipeline, we employed the reference files cataloged under \textsc{jwst\_1235.pmap}, accessible via the CDRS website\footnote{\url{https://jwst-crds.stsci.edu/}}, made available on May 24, 2024. We processed the NIRSpec data using the \textsc{msaexp} v0.8.4 Python package \citep{brammer2022_msaexp}\footnote{\url{https://github.com/gbrammer/msaexp}}. \textsc{msaexp} applies a 1/f correction, identifies snowballs on all the rate files, removes each exposure's biases using a median, and performs a specific noise re-scaling based on empty parts of the exposure. Then \textsc{msaexp} runs parts of the Level 2 {\sl JWST} calibration pipeline and performs a manual background subtraction based on the 2D slit cutouts. For the Level 2 {\sl JWST} calibration pipeline functions, we used the standard Space Telescope Science Institute data reduction pipeline version 1.14.0. The wavelength calibration is based on the NIRSpec instrument model, using a parametric calibration that interpolates between long-slit calibrations \citep{Lutzgendorf2024_spie}. All exposures are incorporated into the final data co-addition. We did not perform any slitloss correction given the complex morphology of the GN~42912 system.  Figure~\ref{fig:oiii5008} shows the final G235H + G395H combined spectrum (top panel). We checked that the final noise array returned by \textsc{msaexp} is consistent with the flux array standard deviation. 

% However, the distinctive two-component morphology of the GN~42912 system poses challenges for background subtraction. The resulting non-compact morphology in the 2D spectra is more extended than other single-component sources observed in the context of the same program. As a result, the standard three-shutter nod pattern for subtracting the overall background in each slit is relatively ineffective, as the source overlaps the central band in each nod observation. To address this issue, we implemented a manual slit-based background extraction method, where the overall background is determined using the median of pixels within a sliding window. Appendix~\ref{app:cleaning} provides details on the implementation of this procedure and the enhancements it enables for the strongest emission lines such as [\ion{O}{iii}] 5008\AA. The final spectrum extraction utilizes a box window covering both emission components.

\begin{table}
	\centering
	\caption{Properties of the two spatially resolved components in GN~42912 observed under {\sl JWST} Project ID: 1871 (PI: Chisholm). We refer to these two components as GN~42912-NE (the one in the northeast) and GN~42912-SW (the one in the southwest). The right ascension and declination are the coordinates of the source in the MSA configuration file. The $z$ values are derived as the median redshift and standard deviation of the H$\beta$, H$\gamma$, [\ion{O}{iii}]~5008~\AA, [\ion{O}{iii}]~4960~\AA, and [\ion{Ne}{iii}]~3869~\AA\ emission lines. The photometry filter values are in nJy. Negative values indicate the flux was not detected. In the lower section of the table, we provide $M_{\rm UV}$, derived based on the F125W filter ($\sim1500$\AA\ at $z\sim7.5$), log($M_{\ast}/M_\odot$), obtained from \textsc{bagpipes} SED fits \citep{carnall2018_bagpipes}, $\beta$, calculated based on the slope of the F125W and F182M filters, and $r_e$, the half-light radius derived from a Sersic profile fit to each component in the F182M filter.  }
	\label{tab:sampprop}
	\begin{tabular}{cccc} % four columns, alignment for each
		\hline
		Property & GN~42912-NE & GN~42912-SW  \\
		\hline
            RA & \multicolumn{2}{c}{12:36:37.91}  \\
            DEC & \multicolumn{2}{c}{+62:18:08.63} \\\relax
		$z$ & 7.50153 $\pm$ 0.00403 &  7.49178 $\pm$ 0.00409   \\\relax
		F182M & 124 $\pm$ 4 & 90 $\pm$ 5  \\\relax
  		F210M & 123 $\pm$ 4  & 84 $\pm$ 6  \\\relax
  		F444W & 266 $\pm$ 6 & 240 $\pm$	8 \\\relax
		F606W & $-2$ $\pm$ 3 & $-6$ $\pm$ 4  \\\relax
  		F775W & 3 $\pm$ 3 &  3 $\pm$ 4 \\
            F814W & $-6$ $\pm$ 2 & 3 $\pm$ 3 \\\relax
            F850LP & 4 $\pm$ 8 &  $-16$ $\pm$ 10  \\
            F105W & 78 $\pm$ 6 &  34 $\pm$ 9 \\
            F125W &  120 $\pm$ 6 & 74 $\pm$ 8 \\
        \hline 
        $M_{\rm UV}$ [mag] &  $-20.89 \pm 0.09$ &  $-20.37 \pm 0.05$ \\
         log$_{10}$($M_{\ast}/M_\odot$) &  8.43$^{+0.04}_{-0.03}$  & 8.87$^{+0.29}_{-0.15}$ \\
         $\beta$ & $-1.92 \pm 0.08$ & $-1.51 \pm 0.16$  \\
         $r_e$ [\arcsec] & 0.026 $\pm$ 0.013 & 0.128 $\pm$ 0.015 \\ 
         \hline
	\end{tabular}
\end{table}

% that because GN~42912-SW is "in front" of GN~42912-NE, its circumgalactic medium (CGM) gas may absorb some of the escaping radiation, including ionizing photons, from GN~42912-NE. However, the indirect diagnostic based on the \ion{Mg}{ii} doublet, as utilized in this study, is unlikely to be significantly affected by this scenario. For GN~42912-SW to impact the escaping \ion{Mg}{ii}, it would require a CGM enriched with \ion{Mg}{ii} capable of absorbing the incoming radiation. Such a configuration is rare and hence unlikely to influence the observed \ion{Mg}{ii} features. \textbf{THIS NEEDS TO BE REMOVED OR EXPLAINED BY SOMEONE WHO can make THIS BETTER}.

\section{Galaxy properties}
\label{sec:meas}
Here we detail the galaxy and emission line properties of the GN~42912 system. Section~\ref{sec:overall} presents an overview and discusses the two-components morphology. Section~\ref{sec:sed} details the SED properties and $\beta$ slopes, and Section~\ref{sec:emission} presents the analysis of the observed emission lines in both galaxies.

\subsection{GN~42912: two luminous companions at $z\sim7.5$}
\label{sec:overall}

GN~42912 was originally reported as a bright ($M_{\rm UV}\sim -21.6$) object at $z\sim 7.5$ \citep{finkelstein2013_lyadet, hutchison2019_ciii,jung2020_lyatexas}. Observations with Keck/MOSFIRE highlighted a significant detection of Ly$\alpha$ (S/N$\sim$10.2) with an equivalent width of 33.2\AA\ \citep{jung2020_lyatexas}. The close correspondence between the mechanisms regulating the escape of ionizing photons and the escape of \Lya\ photons \citep{verhamme2015, verhamme2017,dijkstra2016, gazagnes2020, flury2022_lycdiag} hinted at a potential non-null \fesclyc\ in this system. 

The high spatial resolution of the {\sl JWST} NIRCam and NIRSpec instruments revealed that GN~42912 is composed of two nearby galaxies. We determined the redshift of each object as the median redshift and standard deviation of prominent emission lines fits such as H$\beta$, H$\gamma$, [\ion{O}{iii}] 5008\AA, [\ion{O}{iii}] 4960\AA, and [\ion{Ne}{iii}] 3869\AA\ (see Section~\ref{sec:emission} for the fits details). The northeast component is at a slightly higher redshift compared to the southwest component (7.50153 versus 7.49178, see Table~\ref{tab:sampprop}). The redshift of GN~42912-NE is within $1\sigma$ of the value reported in \citet{hutchison2019_ciii} (7.5032) measured using the \ion{C}{iii}] $\lambda$1909\AA\ emission line.

GN~42912-NE and  GN~42912-SW are separated by 3.1 cMpc assuming Hubble flow, and 358 km s$^{-1}$ in velocity. Figure~\ref{fig:f444w} shows that these two objects are separated by 0$\farcs$1 on the sky, corresponding to $\sim0.5$ kpc. These distances indicate that we cannot rule out non-Hubble flows
or past interactions between them.  

It is worth noting that, despite its significant Ly$\alpha$ detection reported in \citet{jung2020_lyatexas}, Ly$\alpha$ is absent in the PRISM observations of the same object from the JADES program \citep{eisenstein2023_jades,bunker23_jades}. However, the slit position of NIRSpec in these observations is oriented almost perpendicular to the axis passing through the centers of each component, which means that the PRISM observations only partially cover a small portion of each source. Hence, the Ly$\alpha$ emission is likely originating from the non-covered portions in the JADES observations, making it challenging to pinpoint whether it comes from a specific region of one or the other components based on the current data. We discuss further the \Lya\ properties of the GN~42912 system in Section~\ref{sec:bubble} where we infer the size of the ionized region around the two galaxies.

\subsection{SED properties and $\beta$ slopes}
\label{sec:sed}

Utilizing the NIRCam observations, we derive constraints on the SED properties of GN~42912-NE and GN~42912-SW. The F444W imaging, covering wavelengths up to 4.98~$\mu$m (restframe 7500~\AA\ at $z\sim7.5$), is particularly suited for measuring stellar masses ($M_\ast$) because it traces the redder portion of the rest-frame optical spectrum which is more sensitive to older stars. The inference of $M_\ast$ involves fitting the SED using the Bayesian Analysis of Galaxies for Physical Inference and Parameter EStimation (\textsc{bagpipes}, \citealt{carnall2018_bagpipes}) code. \textsc{bagpipes} employs stellar population synthesis templates from \citet{bruzual2003_SPS}, a \citet{kroupa2001_imf} stellar initial mass function, and incorporates nebular emission through the processing of stellar emission via \textsc{cloudy} v17.00 \citep{ferland2017}. We allow for a broad range of stellar masses, metallicities, and ionization parameters, applying log-uniform priors to all three physical properties. $M_\ast$ is determined by constraining the underlying restframe UV to optical continuum, as well as the strength of the prominent [\ion{O}{iii}] and H$\beta$. We fit the data using the same assumptions as in \citet{chisholm24_nev}. The star formation history (SFH) prior follows the "bursty continuity" model from \citet{tacchella2022_sfh}, structured into six distinct time bins. The first two bins cover 0–3 Myr and 3–10 Myr, while the remaining four are logarithmically spaced between 10 Myr and the assumed formation redshift of $z = 20$. The dust attenuation law chosen is the \citet{calzetti2000} law. 
 
With \textsc{bagpipes}, we infer a log($M_{\ast}$/$M_\odot$) of $\sim8.4$ for GN~42912-NE and $\sim8.9$  for GN~42912-SW (Table~\ref{tab:sampprop}). The $M_{\rm UV}$, derived from the F125W filter (restframe 1500~\AA), is $-$20.89 and $-$20.37 for the NE and SW components respectively. Finally, we derive the $\beta$ slopes using a power-law fit to the photometric slope F125W - F182M, which corresponds to the slope between 1500\AA\ and 2200\AA\ at $z\sim7.5$. GN~42912-SW is relatively redder than its NE companion, with a $\beta$ of $-1.51\pm0.16$, compared to $-1.92\pm0.08$. These values are consistent with their $M_{\rm UV}$ as they fall within the trend $M_{\rm UV}$-$\beta$ trend from \citet{bouwens2014_uvcont} which predicts $\beta = -1.77\pm0.19$ for GN~42912-SW and $-1.87\pm0.17$ for GN~42912-NE. However, the two galaxies are less massive than expected from the $\beta$ to stellar mass empirical relations from \citet{finkelstein2012_candels}.

\subsection{Emission lines and Metallicity}
\label{sec:emission}

\newcolumntype{?}{!{\vrule width 1pt}}

\begin{figure*}
    \centering
    \includegraphics[width = \textwidth]{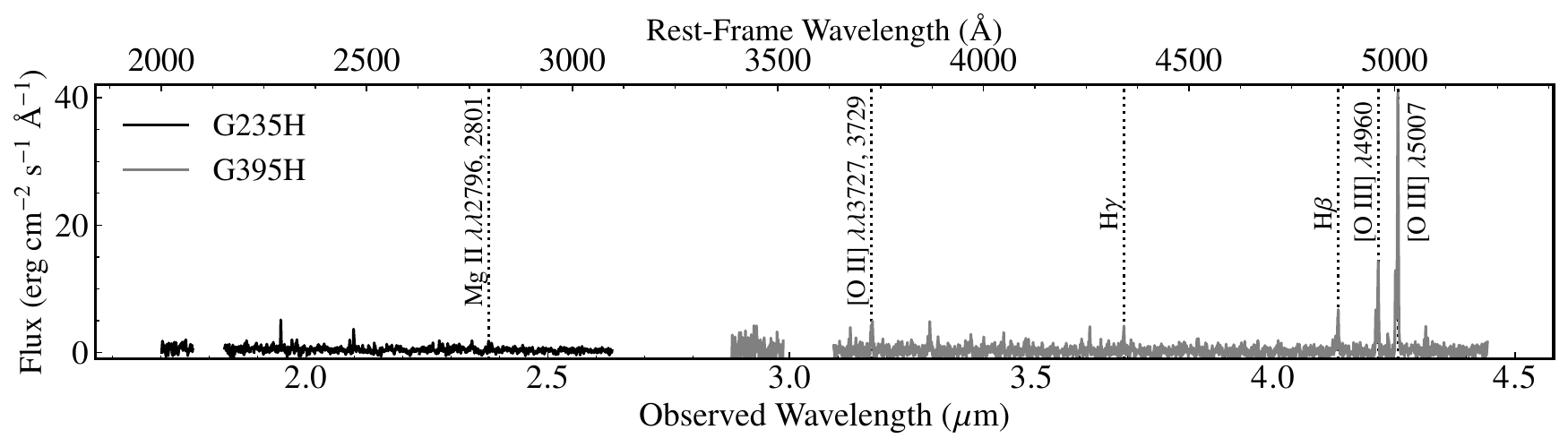} 
    \begin{tabular}{cc?c}
    
\multicolumn{2}{c}{\bf Lines with NE and SW components}   & \quad \bf Lines with NE component only \\
\includegraphics[width = .313\textwidth]{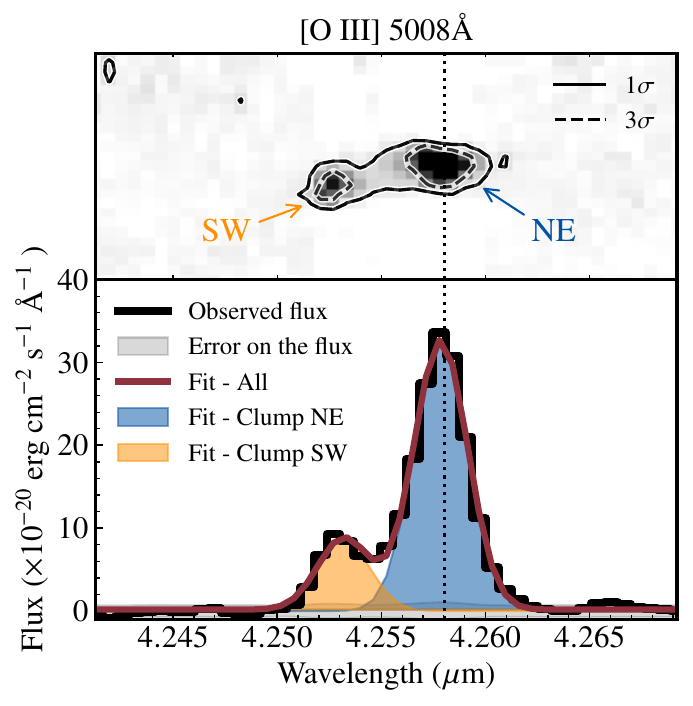} &     \includegraphics[width = .313\textwidth]{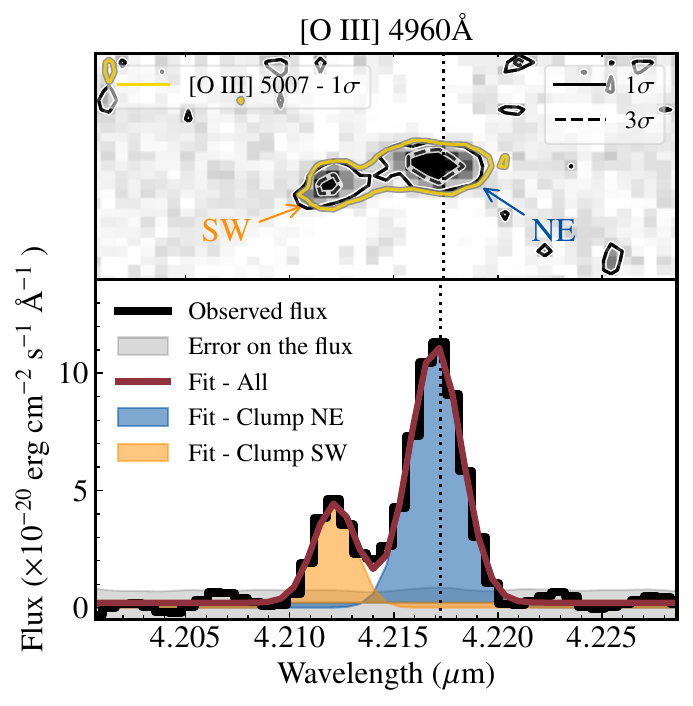} &   \includegraphics[width = .313\textwidth]{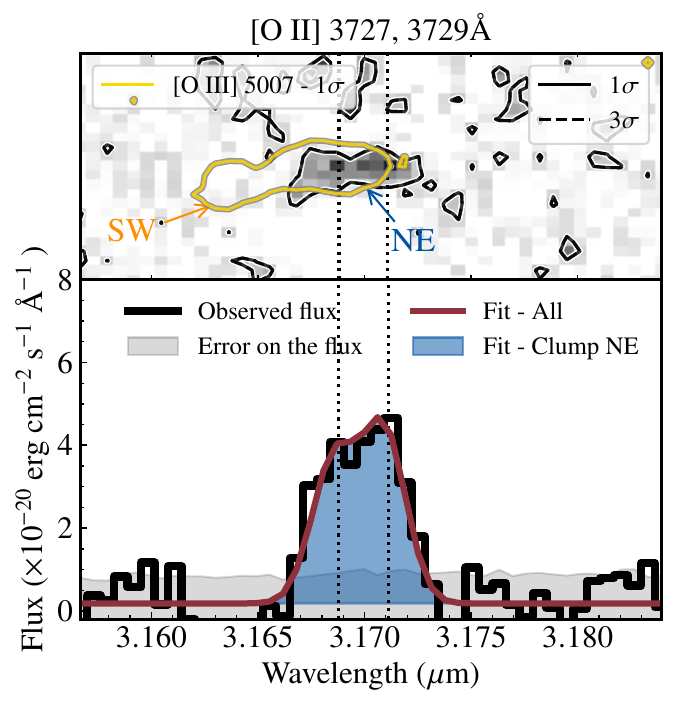} \\
 \includegraphics[width = .31\textwidth]{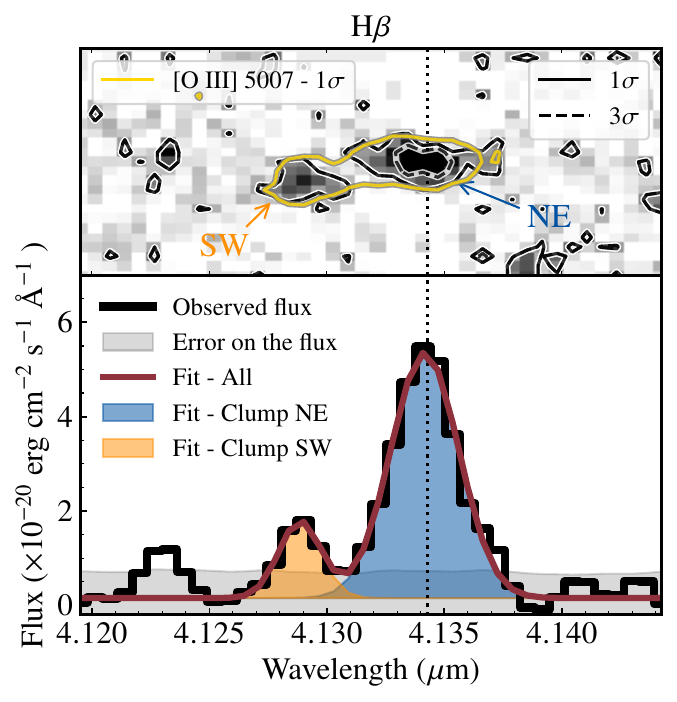}  &   
 \includegraphics[width = .325\textwidth]{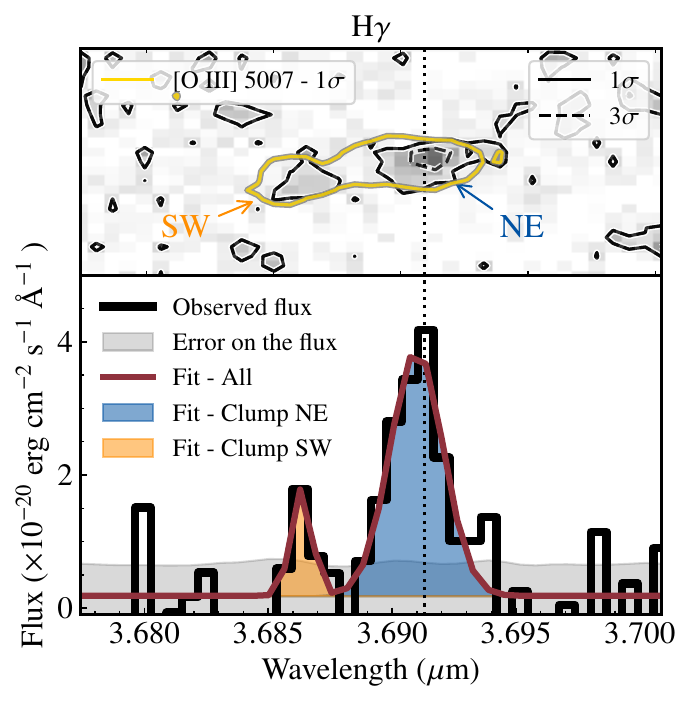}  &    \includegraphics[width = .325\textwidth]{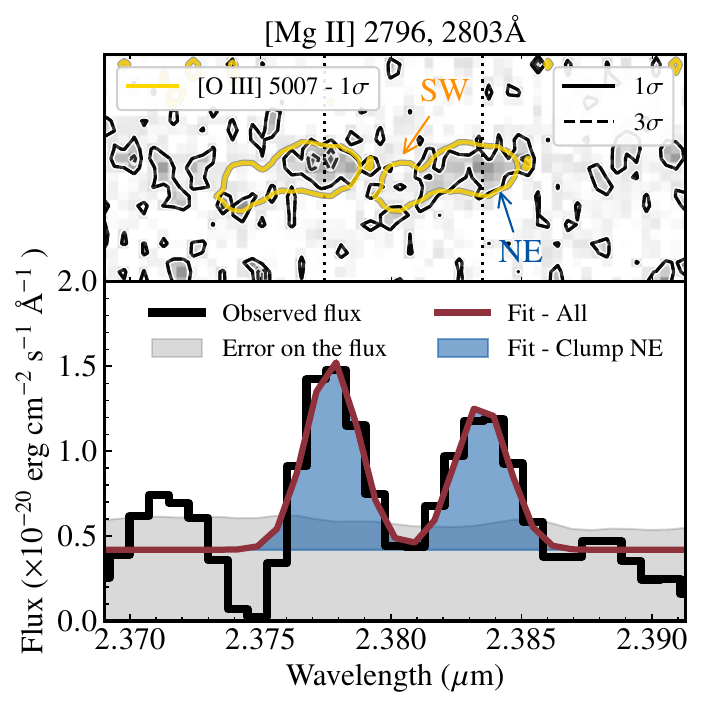}
    \end{tabular}
    \caption{Overview of the spectrum and emission lines of GN~42912-NE and GN~42912-SW. The top panel displays the combined G235H and G395H {\sl JWST} observations for the sum of the spectra of the two NE + SW components. Below, we present each emission line analyzed in this work, categorizing lines detected in both GN~42912-NE and GN~42912-SW (left part) and lines detected only in GN~42912-NE (right part). Each panel features the 2-D spectrum at the top with the 1 and 3 $\sigma$ contours, along with overlaid contours from the [\ion{O}{iii}] 5008\AA\ lines. At the bottom of each panel, observed fluxes are depicted in black, with orange representing the fit to the component of GN~42912-SW, blue representing the fit to the GN~42912-NE component, and red indicating the total fit combining both. Cases where the lines from GN~42912-SW are not detected (rightmost column) include only the fit for the NE component.  }
    \label{fig:oiii5008}
\end{figure*}

Here we outline our methodology for measuring the emission line properties of GN~42912-NE and GN~42912-SW. Our analysis focuses on the [\ion{O}{iii}] 5008\AA, [\ion{O}{iii}] 4960\AA, [\ion{O}{ii}] 3727+3279\AA, H$\beta$, H$\gamma$, and \ion{Mg}{ii} 2796\AA\ and 2803\AA\ emission lines, which will be used to derive indirect constraints on the relative and absolute escape fraction of ionizing photons in Section~\ref{sec:mgII}. These lines are all detected at a significance level of $>3\sigma$ for the NE component. However, in the SW component which mostly falls outside the slit, we do not detect [\ion{O}{ii}] 3727+3729\AA\ nor \ion{Mg}{ii} 2796+2803\AA. Figure~\ref{fig:oiii5008} bottom panels show the 2D and 1D extracted spectra for each of the lines of interest, distinguishing between lines detected in both galaxies and those detected only in GN~42912-NE.

We employ a single Gaussian profile to fit each resolved line and use a double Gaussian to fit the blended [\ion{O}{ii}] 3727+3279\AA\ lines. A separate paper will delve into a more sophisticated fitting approach to analyze potential contributions from a broad component (Saldana-Lopez, in prep). For our current analysis, we find that using a single Gaussian profile is sufficiently accurate to replicate the observed profiles. The velocity widths of lines from the same ion are tied together. The fits are displayed in Figure~\ref{fig:oiii5008}. In cases where we do not detect emission lines in GN~42912-SW, we establish an upper limit on the fluxes of these lines by integrating the error spectrum across the same line width observed for the NE component ($\sim$100 km s$^{-1}$). The integrated flux measurements are reported in Table~\ref{tab:ELprop}.

We calculate the rest-frame equivalent widths $W$ for all lines by using a continuum taken as the median flux within a $\pm$ 5,000 km\,s$^{-1}$ feature-free interval on both sides of the lines. Since we cannot disentangle the contributions of each object in the observed continuum of the final spectra, we adjust the continuum values by the ratio of the photometric values, taking the ratio over the {\sl JWST} filter falling closest to the line. Specifically, for the NE component, the observed continuum is multiplied by a factor of 0.525 for lines falling close to the F444W filter and of 0.594 for lines falling close to the F210M filter.

The emission line measurements for GN~42912-NE and GN~42912-SW are reported in Table~\ref{tab:ELprop}. The latter table also presents some relevant line ratios for the present work. The O$_{32}$ ratio, defined as [\ion{O}{iii}] 5008\AA/[\ion{O}{ii}]~3727+3279\AA, is 5.3$\pm$0.7 for GN~42912-NE and a lower limit of 5.8 for GN~42912-SW. The  H$\gamma$/H$\beta$ Balmer line ratio provides insights into the dust extinction in both objects. In GN~42912-SW, we find H$\gamma$/H$\beta$ ratio of 0.462$\pm0.167$, close to the theoretical expectation of $\sim0.468$ for an \ion{H}{ii} region under Case B recombination at $T_e$ = 10,000K and $n_e=500$ cm$^{-3}$ \citep{Osterbrock2006}. This ratio indicates a relatively dust-free nebular environment, which contrasts with the relatively dusty continuum suggested by the photometry with $\beta\sim-1.5$. In GN~42912-NE, we find a Balmer H$\gamma$/H$\beta$ ratio of 0.523$\pm$0.087. This value is larger than the theoretical dust-free value. The relatively significant errors on these line ratios prevent us from concluding whether such deviations from the theoretical predictions are true or are the result of statistical errors in the flux determination. Furthermore, as shown in Figure~\ref{fig:f444w}, the JWST observations capture only part of the galaxy system, likely including a fraction of the \ion{H}{ii} regions but not the entire structure. High-redshift galaxies, particularly massive ones, often exhibit a clumpy morphology, leading to inhomogeneous dust distribution. Consequently, dust attenuation inferred from global photometry may not accurately represent the conditions within the captured \ion{H}{ii} regions.

% Balmer line ratios exceeding the dust-free theoretical value have recently been identified in high-$z$ objects. \citet{topping2024metalpoor} found notably higher H$\gamma$/H$\beta$ and H$\delta$/H$\beta$ values in RXCJ2248-ID at $z = 6.11$. \citet{cameron2024nebular} noted a H$\alpha$/H$\beta$ ratio significantly ($>1\sigma$) smaller than expected without dust extinction in GS-NDG-9422, a $z\sim5.943$ galaxy. Recent studies have proposed different scenarios to elucidate these ``anomalies'' \citep{yanagisawa2024balmer, scarlata2024universal}. These scenarios involve unique environments that depart from the typical Case B assumptions and may include Balmer self-absorption (see \citealt{scarlata2024universal} or \citealt{Osterbrock2006}), or collisionally excited \ion{H}{i} clouds \citep{yanagisawa2024balmer}. Nevertheless, understanding the origin of Balmer ``anomalies'', whether it is due to flux errors or a specific physical regime, is important for accurately determining the characteristics of these galaxies, yet this effort lies beyond the scope of our present analysis. 

In the context of our study, Balmer decrements are important for correcting the influence of dust on the nebular emission lines. However, given the lack of reliability of the measured H$\gamma$/H$\beta$ ratios,  we do not use these ratios to correct for the dust attenuation. Instead, we adopt two distinct approaches for our analysis. In the first scenario, we assume a dust-free case, thus considering the observed flux values as intrinsic (hereafter, this case is referred to as no-dust-case or NDC). Alternatively, in the second scenario, we correct for dust in the nebular emission lines using the $\beta$ values as proxies for the colour excess $E(B-V)$ (hereafter, this case is referred to as $\beta$-dust-case or $\beta$DC). 
To estimate $E(B-V)$ from $\beta$, we use equation 7 from \citet{chisholm2022_beta}. Because the shape of the dust extinction is unconstrained for reionization-era galaxies, we use both the SMC dust attenuation law \citep{gordon2003} and the dust law proposed by \citet{reddy2016dustlaw}, derived from a large dataset of 933 far-UV observations of Lyman Break Galaxies at $z\sim3$. 

Using both the NDC and $\beta$DC dust scenarios, along with two different dust extinction laws, allows for a comprehensive investigation into the impact of dust on the derived \fesclyc\ for both objects. However, this approach does not imply that all scenarios considered are equally plausible. The measured $\beta$ slopes for both objects indicate that the escaping radiation is attenuated by dust, suggesting that the NDC scenario may represent an extreme case. Moreover, $\beta$ reflects the dust attenuation along the stellar line of sight, while nebular dust extinction is generally greater (E(B-V)$_{\rm stellar}$ = 0.44E(B-V)$_{\rm nebular}$; \citealt{calzetti2000}). Consequently, the $\beta$DC scenario may underestimate the true level of nebular dust extinction. In Section~\ref{sec:dust}, we further explore how these assumptions affect the final \fesclyc\ estimates, demonstrating that a higher nebular dust extinction than considered here would lead to lower \fesclyc\ estimates and, therefore, would not alter the overall conclusions of this study.

\subsubsection{Metallicity}

Here, we estimate the O/H abundance ratio in GN~42912-NE and GN~42912-SW (needed to estimate the intrinsic \ion{Mg}{ii} emission and calculate \fesclyc\ below). We combine two direct empirical metallicity calibrations utilizing the R$_{23}$ ratio (defined below) and the O$_{32}$ ratio from \citet{curti2020_gasphase}. The relation between R$_{23}$ or O$_{32}$ and 12+log(O/H) is given as

\begin{align}
    \text{log}({\rm R}_{23}) &= 0.527 - 1.569\times x - 1.652\times x^2 -0.421\times x^3 \\
    \text{log}({\rm O}_{32}) &= 0.691 - 2.944\times x - 1.308\times x^2 \\ 
    \text{with}\ x &= 12+\text{log(O/H)} - 8.69\ \text{, }\ \nonumber\\
     {\rm R}_{23} &= \frac{I([\ion{O}{ii}]\ \lambda\lambda3727,3729 + [\ion{O}{iii}]\ \lambda5008 +[\ion{O}{iii}]\ \lambda4960)}{I({\rm H}\beta)} \nonumber\\
     \text{ and }\ {\rm O}_{32} &= \frac{I([\ion{O}{iii}]\ \lambda5008)}{I([\ion{O}{ii}]\ \lambda\lambda3727,3729)}.
\end{align}

\noindent$I$ represents the dust-attenuation-corrected flux values of each line. The approach we used relies on finding the log(O/H) value that minimizes both differences to the measured R$_{23}$ or O$_{32}$ based on the empirical relations found in \citet{curti2020_gasphase}. We note that we have also applied the same strategy using the empirical relations from \citet{sanders2024_jwst} which presents relations based on high-$z$ observations and found consistent estimates.

For GN~42912-NE, we derive ${\rm R}_{23}=$ 8.4$\pm$1.3 and 12+log(O/H) = 7.93$\pm$0.11 in the NDC case, and ${\rm R}_{23}=$ 8.5$\pm$1.4 and 12+log(O/H) = 7.95$\pm$0.16 in the $\beta$DC. The absence of significant changes is expected because the [\ion{O}{ii}] emission is sufficiently weak that the dust correction has minimal impact on the results. Given the consistent central value, we adopt the value calculated with the $\beta$DC which is more conservative. %, as nebular dust attenuation is generally greater than stellar dust attenuation \citep{calzetti2000}.

For GN~42912-SW, the [\ion{O}{ii}] $\lambda\lambda$3727,3729 doublet is not detected, but we find that using the reported upper limit or assuming $F$([\ion{O}{ii}])~=~0 only marginally influences the final 12+log(O/H) estimate by 0.01. Further, similar to GN~42912-NE, using either the NDC or the $\beta$DC case does not influence the final 12+log(O/H) estimate. We find ${\rm R}_{23}=$ 11.8 $\pm$ 3.3, resulting in 12+log(O/H) $=$ 8.05 $\pm$ 0.24. \citet{finkelstein2013_lyadet} reported with 95 per cent confidence that the gas-phase metallicity in this system (not resolved in their observations) was sub-solar, with a stellar metallicity between 20 per cent and 40 per cent based on SED fitting. This is consistent with our estimates, where GN~42912-NE is 18 per cent of Z$_\odot$ and GN~42912-SW is 23 per cent of Z$_\odot$.

In the next section, we detail the two \ion{Mg}{ii}-based strategies to derive an estimate and an upper limit on the escape fraction of ionizing photons in GN~42912-NE and GN~42912-SW, respectively.

\begin{table*}
	\centering
	\caption{Emission line properties in GN 42912. Fluxes ($F_\lambda$) are in 10$^{-20}$ erg cm$^{-2}$ s$^{-1}$ and have not been corrected for dust attenuation. The rest-frame equivalent widths ($W_\lambda$) are in \AA. The lower section of the table provides properties derived using the line measurements. }
	\label{tab:ELprop}
	\begin{tabular}{ccccc} % four columns, alignment for each
		\hline
		Properties & \multicolumn{2}{c}{GN~42912-NE} & \multicolumn{2}{c}{GN~42912-SW}  \\
   & $F_\lambda$ & $W_\lambda$ & $F_\lambda$ & $W_\lambda$ \\
		\hline
		[\ion{O}{iii}] $\lambda$5008 & 1052 $\pm$ 26 & 1090 $\pm$ 31   & 271 $\pm$ 22 
 &  312 $\pm$ 35  \\\relax
		[\ion{O}{iii}] $\lambda$4960 & 354 $\pm$ 20  & 329 $\pm$ 27  & 109 $\pm$ 24 & 124 $\pm$ 32  \\\relax
            H$\beta$    & 188 $\pm$ 14  & 193 $\pm$ 32 &    37 $\pm$ 9 & 42 $\pm$ 26 \\\relax
            H$\gamma$  & 99 $\pm$ 17   & 113 $\pm$ 27 & 17 $\pm$ 9 & 20 $\pm$ 18  \\\relax
  		[\ion{O}{ii}] $\lambda$3727  & 89 $\pm$ 16  & 80 $\pm$ 28  & $<23$  & -\\\relax
  		[\ion{O}{ii}] $\lambda$3729  & 108 $\pm$ 17   & 96 $\pm$ 27 & $<23$ & -  \\\relax
		\ion{Mg}{ii} $\lambda$2796 & 28 $\pm$ 4  & 14 $\pm$ 4  & $<10$ & -  \\\relax
  		\ion{Mg}{ii} $\lambda$2803 & 22 $\pm$ 4 & 11 $\pm$ 4    & $<10$ & - \\

        \hline 
    R$_{\ion{Mg}{ii}}$ & \multicolumn{2}{c}{1.28 $\pm$ 0.26} & \multicolumn{2}{c}{-}  \\
    O$_{32}$ & \multicolumn{2}{c}{5.3 $\pm$ 0.7} & \multicolumn{2}{c}{$>$ 5.8}  \\
    H$\gamma$/H$\beta$ & \multicolumn{2}{c}{0.523 $\pm$ 0.087} & \multicolumn{2}{c}{0.462 $\pm$ 0.167} \\
    R$_{23}$ & \multicolumn{2}{c}{8.4 $\pm$ 1.3} &  \multicolumn{2}{c}{11.8 $\pm$ 3.3} \\
    12+log(O/H)$_{R_{23}}$ & \multicolumn{2}{c}{7.95 $\pm$ 0.16} & \multicolumn{2}{c}{8.05 $\pm$ 0.28}  \\
		\hline
	\end{tabular}
\end{table*}

\section{Mg II Escape Fractions}
\label{sec:mgII}

In this section, we outline our methodology for deriving indirect \fesclyc\ constraints utilizing the \ion{Mg}{ii} $\lambda\lambda$2796,2803\AA\  doublet lines. The \ion{Mg}{ii} doublet has been suggested as a potential indirect indicator of LyC escape fraction \citep{henry2018}, owing to its low ionization potential which is close to that of hydrogen. This characteristic makes it a valuable tracer of neutral gas and more easily observed at higher redshifts compared to Ly$\alpha$, which is directly affected by the optically thick \ion{H}{i} IGM at $z>6$. While a recent study from \citet{katz2022_mgii} has noted important caveats to using \ion{Mg}{ii} as a LyC tracer from an optically thick medium (these concerns are discussed in Section~\ref{sec:doublet_cav}), observational studies have demonstrated a close correspondence between direct \fesclyc\ measurements at $\sim$ 912\AA\ and indirect estimates based on the \ion{Mg}{ii} 2796,2803\AA\ emission lines \citep{chisholm2020, xu2023_lycmgII, leclercq2024_lycMgII}. In this study, we explore two approaches for calculating \fesclyc\ from these \ion{Mg}{ii} lines: using the doublet ratio method (Section~\ref{sec:doublet}) and using photoionization models (Section~\ref{sec:photo}). 

\subsection{\ion{Mg}{ii} Doublet ratio method}
\label{sec:doublet}

The \ion{Mg}{ii} emission line ratio depends on their emissivities, where the emissivity relates to the Einstein A coefficient of the \ion{Mg}{ii} transitions and on the number density of electrons populating the upper levels. When collisions dominate the Mg$^+$ excitation, the intrinsic flux ratio of the \ion{Mg}{ii} 2796\AA\ and 2803\AA\ emission lines (hereafter R(\ion{Mg}{ii})) is $\approx2$. This value is confirmed by \textsc{cloudy} photoionization modeling \citep{henry2018} and Monte Carlo radiative transfer simulations \citep{chang2024_mgii}. In the scenario where the excitation is dominated instead by resonant scattering and photon absorption, the emission flux ratio is dominated by the ratio of the Einstein A coefficients. Both \ion{Mg}{ii} 2796\AA\ and 2803\AA\ lines have similar A$_{21}$ values, so  R(\ion{Mg}{ii}) would come closer to 1 instead of 2. 

\citet{chisholm2020} showed that, in the optically thin regime, one can use the variation of the R(\ion{Mg}{ii}) values to infer the neutral hydrogen column density, which in turn provides a constraint on the relative absorption of the LyC photons by the H$^0$ gas. Such an approach is possible because (1) in the optically thin regime, resonant radiative transfer effects on the \ion{Mg}{ii} lines are negligible and (2) the Mg$^0$ and Mg$^+$ phases overlap with H$^0$, so a column density of  Mg$^+$ can be transformed into  H$^0$ under some assumption of the Mg/H abundance in the galaxy. 

A robust Mg$^+$ to H$^0$ conversion requires accurate estimates of the depletion fraction of Mg and of the Mg/H abundance ratio. Accurately determining the fraction of Mg that is depleted into dust is complex, and even more complex for reionization-era galaxies where the number of studies on the topic remains scarce. For consistency with previous studies using a similar approach, we assume a depletion fraction of 27 per cent based on \citet{jenkins2009}, derived from Milky Way observations. As noted in \citet{chisholm2020}, there exists a substantial scatter in the distribution of measurements of the depletion factor, yet, this factor should not appreciably vary with metallicity for star-forming galaxies \citep{guseva2013,guseva2019}. Regarding the Mg/H abundance ratio, since both oxygen and magnesium are $\alpha$ elements primarily produced by core-collapse supernovae, the Mg/O value should not appreciably vary \citep{guseva2019}. This means that we can approximate the Mg/H abundance ratio using the observed O/H abundance ratio and make the conversion assuming a solar O/Mg abundance ratio of 12.3 \citep{asplund2021}. 

Under these assumptions, \citet{chisholm2020} showed that the \ion{H}{i} column density can then be derived as

\begin{equation}
\label{eq:nh}
    N_{\rm \ion{H}{i}} = -2\times 10^{13} {\rm cm}^{-2}\ \frac{\rm H}{\rm O}\ {\rm ln}({\rm R(\ion{Mg}{ii})}/2).
\end{equation}

\noindent From the estimate of $N_{\rm \ion{H}{i}}$, we can derive a relative escape fraction, $f^{\rm Lyc}_{\rm esc, rel}$(LyC) as

\begin{equation}
\label{eq:fescrel}
    f^{\rm LyC}_{\rm esc, rel} = e^{-N_{\rm \ion{H}{i}}\times\sigma_{\nu0}},
\end{equation}

\noindent where $\sigma_{\nu 0}$ is the ionisation cross-section of hydrogen (6.3$\times$10$^{-18}$\pcm). $f^{\rm LyC}_{\rm esc, rel}$ represents the amount of escaping LyC photons solely based on the quantity of neutral hydrogen and disregarding the influence of dust within the galaxy.
To correct for the ionizing photons absorbed by dust, we calculate the absolute escape fraction $f^{\rm LyC}_{\rm esc, abs}$ accounting for the reduction in the relative escape fraction due to the expected dust attenuation at 912\AA:
\begin{equation}
\label{eq:fescabs}
    f^{\rm LyC}_{\rm esc, abs} = f^{\rm LyC}_{\rm esc, rel}\times 10^{-0.4E(B-V)k(912\text{\AA})},
\end{equation}
where $E(B-V)$ is the line-of-sight colour excess and $k(912\text{\AA})$ is the predicted dust attenuation from the chosen dust attenuation law at 912\AA. We note that direct and indirect measurements of the absolute \fesclyc\ established from observations are line-of-sight dependent, as these observations only allow us to trace the photons escaping towards us. In practice, the overall (i.e., angle-averaged) \fesclyc\ is the quantity that is relevant for cosmological reionization models. Since the LyC escape is highly anisotropic \citep{gazagnes2020, saldana2022, mauerhofer2021, flury2022_lycdiag, Choustikov2024_lyc}, transforming line-of-sight constraints into a global constraint can be done using the average of line-of-sight-dependent \fesclyc\ constraints over a statistically significant sample of galaxies. While more high redshift observations are needed to build these samples, the \fesclyc\ constraints set in the work may be considered a first step toward this goal.

The doublet ratio method is only applicable to galaxies that have detections of the \ion{Mg}{ii} 2796 and 2803 \AA\ lines. Hence, in this work, we only apply this method to GN~42912-NE. We derive $N_{\rm \ion{H}{i}}$ using Equation~\ref{eq:nh}, taking the O/H abundance measured from the R$_{23}$ ratio (Table~\ref{tab:ELprop}). We obtain a $N_{\rm \ion{H}{i}}$ of $(1.04 \pm 0.55) \times 10^{17}$ \pcm. We find that using either the NDC or $\beta$DC scenarios yields similar values.

Using Eq~\ref{eq:fescrel}, we find that the measured  $N_{\rm \ion{H}{i}}$ value corresponds to $f^{\rm LyC}_{\rm esc, rel}$ of $0.52 \pm 0.18$. Physically, this means that the neutral hydrogen absorbs approximately 48 per cent of the escaping LyC radiation. As a comparison, $z\sim0$ galaxies with a significant detection of the LyC leakage tend to have $f^{\rm LyC}_{\rm esc,\ rel}$ closer to 100 per cent (see \citealt{chisholm2022_beta} and further comparisons in Section~\ref{sec:robustness}). 

In the NDC scenario, where we assume that the galaxy is completely devoid of dust, $f^{\rm LyC}_{\rm esc, rel}$ is also $f^{\rm LyC}_{\rm esc, abs}$ since no attenuation is expected below 912\AA. In the $\beta$DC scenario, we derive $f^{\rm LyC}_{\rm esc, abs}$ using Equation~\ref{eq:fescabs} and $E(B-V)$ values transformed from $\beta$ using Equation~7 from \citet{chisholm2022_beta}. To investigate the impact of the dust attenuation law choice, we utilize both the SMC attenuation curve \citep{gordon2003} and the attenuation curve from \citet{reddy2016dustlaw} (hereafter R16). We selected these two laws because their attenuation curves differ significantly in the UV, enabling us to compare the variations in $f^{\rm LyC}_{\rm esc}$ resulting from this choice. We derive an $f^{\rm LyC}_{\rm esc, abs}$(R16) of $0.105 \pm 0.040$ and $f^{LyC}_{\rm esc, abs}$(SMC) of $0.142 \pm 0.052$. The consistency between the two estimates, both within $1\sigma$ of each other, indicates that the choice of the dust attenuation law has little impact on the constraint on the LyC leakage in GN~42912-NE.

The doublet ratio method comes with a notable caveat: its reliability strongly depends on the validity of the optically thin regime where the ratio effectively traces $N_{\rm \ion{H}{i}}$. However, when the observed ratio deviates from this optically thin scenario (i.e., R(\ion{Mg}{ii}) closer to 1), resonant radiative transfer effects can become non-negligible, introducing a bias in the relationship from R(\ion{Mg}{ii}) to $N_{\rm \ion{H}{i}}$. R(${\ion{Mg}{ii}}$) is 1.28 in GN~42912-NE, suggesting that the optically thin assumption may not be valid in this case. This caveat suggests that we should view the \fesclyc\ derived with the doublet ratio method as an upper limit rather than a well-defined measurement.  We discuss this aspect further in Section~\ref{sec:cavdoub}.

\begin{table*}
	\centering
	\caption{\ion{Mg}{ii} properties and \ion{Mg}{ii}-based LyC escape fractions for GN~42912-NE and GN~42912-SW. The Doublet Ratio method refers to the \citet{chisholm2020} method for determining \fesclyc\ from the \ion{Mg}{ii} doublet and the Photoionization models method was originally introduced by \citet{henry2018}. We use two different scenarios, a dust-free case (NDC) and a $\beta$ dust-case ($\beta$DC), where we use $\beta$ as a tracer of both nebular and stellar E(B-V). We also consider two dust extinction laws, the SMC dust extinction law \citep{gordon2003} and the \citet{reddy2016dustlaw} dust extinction law. Both \ion{Mg}{ii} and [\ion{O}{ii}] are not detected in GN~42912-SW, so we adopt an upper limit on the relative escape fraction of 100\%. The upper limit on $f^{\rm LyC}_{\rm esc, abs}$ is derived by correcting the relative escape fractions using the dust attenuation at 912\AA. The bottom part of the table shows the final \fesclyc\ estimates that we use when comparing our constraints to low-$z$ LyC trends in Section~\ref{sec:lyclit} and reionization models in Section~\ref{sec:reio}. These estimates correspond to the $\beta$DC and use the SMC dust extinction law. The reasoning behind this choice is detailed in Section~\ref{sec:robustness}. }
	\label{tab:mgiitab}
	\begin{tabular}{ccccc} % four columns, alignment for each
		\hline
		Property &   \multicolumn{2}{c}{GN~42912-NE} & \multicolumn{2}{c}{GN~42912-SW}\\ 
   & NDC & $\beta$DC &NDC & $\beta$DC  \\ \hline
    %             EW($\lambda$2796) &  9.9 $\pm$ 2.2 & -  \\\relax
    % EW($\lambda$2803) & 10.7 $\pm$ 2.6 & - \\
            \multicolumn{5}{l}{\bf \ion{Mg}{ii} Doublet Ratio method ------------------------ } \\
             % R$_{23}$ & 7.1 $\pm$ 1.1 & - \\
             % 12+log(O/H)$_{R_{23}}$  & 7.68 $\pm$ 0.19 & -  \\
             R$_{\ion{Mg}{ii}}$ & \multicolumn{2}{c}{1.28 $\pm$ 0.26} & \multicolumn{2}{c}{-} \\
             N$_{\rm H_0}$ &  \multicolumn{2}{c}{(1.05 $\pm$ 0.55) $\times 10^{17}$ cm$^{-2}$} & \multicolumn{2}{c}{-}   \\
             $f_{\rm esc, rel}^{\rm LyC}$ & \multicolumn{2}{c}{0.518 $\pm$ 0.176} & \multicolumn{2}{c}{-}  \\
             \multirow{ 2}{*}{$f_{\rm esc}^{\rm abs}$(LyC)} & \multirow{2}{*}{0.518 $\pm$ 0.176} & 0.105 $\pm$ 0.040 (R16)& \multicolumn{2}{c}{-} \\
              & & 0.142 $\pm$ 0.052 (SMC) & \multicolumn{2}{c}{-} \\            
            \multicolumn{5}{l}{\bf \ion{Mg}{ii} Photoionization models method ------------------------} \\
            R$_{2796}$  &  $-1.13\ \pm\ 0.10$ &  $-1.10\ \pm\ 0.10$ & $<-1.18$ & $<-1.14$ \\
            % R$_{2803}$  & $-1.50\ \pm\ 0.12$  & $-2.15 \pm 0.10$ \\
		$f_{\rm esc,\ rel}^{\rm \ion{Mg}{ii}\ 2796}$ & 0.359 $\pm$ 0.089   & 0.310 $\pm$ 0.079 & $\leq 1.000$ & $\leq 1.000$ \\ 
		% $f_{\rm esc}^{\rm rel}$(2803) & 0.491 $\pm$ 0.233   & $\leq$ 100\% \\ 
		\multirow{ 2}{*}{$f_{\rm esc, abs}^{\rm LyC}$} &  \multirow{ 2}{*}{0.359 $\pm$ 0.089} & 0.063 $\pm$ 0.019  (R16) & \multirow{ 2}{*}{$\leq 1.000$} & $<0.089$ (R16)   \\ 
  & & 0.085 $\pm$ 0.022 (SMC) & & $<0.140$ (SMC) \\
		% \multirow{ 2}{*}{$f_{\rm esc}^{\rm abs}$(2803 $\rightarrow$ LyC)} &  0.096 $\pm$ 0.071 (R16) & $<0.09$\\ 
  % & 0.131 $\pm$ 0.096 (SMC) & $<0.15$\\
  \multicolumn{5}{l}{\bf Final estimates ------------------------} \\
        $f_{\rm esc,\ rel}^{\rm LyC}$ &  \multicolumn{2}{c}{$<0.310$} &  \multicolumn{2}{c}{$\leq1.000$} \\
       $f_{\rm esc,\ abs}^{\rm LyC}$ &  \multicolumn{2}{c}{$<0.085$} &  \multicolumn{2}{c}{$<0.140$} \\

		\hline
	\end{tabular}
\end{table*}

\subsection{\ion{Mg}{ii} Photoionization method}
\label{sec:photo}

Using \textsc{cloudy} photoionization models, \citet{henry2018} showed that there exists a tight correlation between the intrinsic flux of \ion{Mg}{ii} and the extinction-corrected flux of [\ion{O}{iii}] 5008 \AA\ and [\ion{O}{ii}] 3727,3729 \AA, offering an alternative approach to indirectly deriving \fesclyc. \citet{henry2018} fitted a quadratic equation that relates the intrinsic flux ratio of \ion{Mg}{ii}\,$\lambda$2796/[\ion{O}{iii}] $\lambda$5008 and the O$_{32}$ ratio. \citet{xu2022_lya_mgII} extended the work of \citet{henry2018} by deriving the coefficient of this equation for three different gas-phase metallicities log($Z/Z_\odot$) of $-1.5$, $-1$, and 0.5, and two geometries, the ionization bounded geometry, where most clouds are neutral and optically thick to escaping photons, and the density bounded geometry, where most clouds are optically thin. Here, we adopt the equation with gas-phase metallicity closer to the derived 12+log(O/H) values (i.e. 18 per cent solar for GN~42912-NE and 23 per cent solar for GN~42912-SW). Given the moderate O$_{32}$ ratio observed in GN~42912-NE (5.3$\pm$0.7), we adopt the coefficients derived under the ionization-bounded scenario and log($Z/Z_\odot$) = $-1$ such that the intrinsic flux ratio of \ion{Mg}{ii} $\lambda$2796/[\ion{O}{iii}] $\lambda$5008 from \citet{xu2022_lya_mgII} is expressed as follows: 

\begin{align} 
    R_{2796} &=  0.074\times x^2\ - 0.97\times x\ - 0.46\ \label{eq:r2796}  \\
    \text{with}\ R_{2796} &= \text{log(}I({\ion{Mg}{ii}\ \lambda2796})/I({[\ion{O}{iii}]\ \lambda5008}){\rm )}\ \text{and} \nonumber \\
    x &= {\rm log(O_{32})}, \nonumber
\end{align}

\noindent where O$_{32}$ denotes the dust-free (NDC) and extinction-corrected ($\beta$DC) O$_{32}$ ratio depending on the case considered. We also explored using the coefficients corresponding to the density-bounded scenario case and for a different gas-phase metallicity. We found a maximum variation of 10 per cent in our estimates of the relative escape fractions, which corresponds to the typical uncertainty on this measurement.

We use Equation~\ref{eq:r2796} and the dust-corrected O$_{32}$ and [\ion{O}{iii}] emission line fluxes to derive the intrinsic \ion{Mg}{ii} 2796 \AA\ line flux (the strongest line of the doublet) for GN~42912-NE and GN~42912-SW. We compare these estimates with the dust-corrected \ion{Mg}{ii} line fluxes to deduce the relative escape fractions at 2796 \AA\ ($f_{\rm esc, rel}^{\rm \ion{Mg}{ii}\ 2796}$), considering both the NDC and the $\beta$DC scenarios. For GN~42912-NE, we derive $f_{\rm esc, rel}^{\rm \ion{Mg}{ii}\ 2796}$ = 0.369 $\pm$ 0.089 in the NDC and 0.310 $\pm$ 0.079 in the $\beta$DC. For the NDC, this relative escape fraction directly translates to the absolute LyC escape fraction because we have assumed that there is no dust in the system. For the $\beta$DC, we convert the relative escape fraction into an absolute \fesclyc\ by multiplying it by the dust attenuation at 912 \AA, employing $E(B-V)$ values extrapolated from $\beta$ \citep{chisholm2022_beta}. We find $f_{\rm esc,\ abs}^{\rm LyC}$ to be 0.063 $\pm$ 0.019 using the R16 dust attenuation law and 0.085 $\pm$ 0.022 using the SMC law. These values are consistent at 1$\sigma$ with the values derived using the doublet ratio method (Section~\ref{sec:doublet}).

The absence of \ion{Mg}{ii} and [\ion{O}{ii}] detections prevent us from applying the same approach to GN~42912-SW. This is because the current lower limit on  O$_{32}$ can only set an upper limit on the intrinsic \ion{Mg}{ii} flux. Since the \ion{Mg}{ii} 2796 \AA\ line flux is also an upper limit, we cannot constrain a reliable estimate of $f_{\rm esc, rel}^{\rm \ion{Mg}{ii}\ 2796}$. Hereafter, we use $f_{\rm esc, rel}^{\rm \ion{Mg}{ii}\ 2796}$ $=1$, which is the most conservative limit we can set, as it assumes all the intrinsically \ion{Mg}{ii} photons are escaping in the absence of dust. With this assumption and using the expected dust attenuation at 912 \AA, we derive $f_{\rm esc,\ abs}^{\rm LyC}$  $\leq0.089$ using the R16 law and $\leq0.140$ using the SMC dust extinction law. 

 Table~\ref{tab:mgiitab} summarizes all the \fesclyc\ estimates derived in this section. Overall, for GN~42912-NE, the photoionization model approach to constraining \fesclyc\ yields results very consistent with the doublet ratio method. For GN~42912-SW, due to the absence of [\ion{O}{ii}] and \ion{Mg}{ii} detections, \fesclyc\ can range from 0 to 100 per cent in the NDC and between 0 and 14 per cent in the $\beta$DC. As mentioned in Section~\ref{sec:emission}, we chose to consider different dust scenarios to explore their implications for \fesclyc, but further discussion is warranted regarding the plausibility of each scenario. Additionally, \ion{Mg}{ii}-based approaches to estimating \fesclyc\ also suffer caveats that we must carefully consider. The next section will focus on this discussion, aiming at establishing a final \fesclyc\ estimate with greater confidence for GN~42912-NE and GN~42912-SW.

% Overall, the significant dust extinction present in GN~42912-SW is likely the largest sink for the escape of ionizing photons. Irrelevant of the actual relative escape fractions, dust should remove more than 85\% of the escape ionizing radiation, which is significant considering that in an oligarch type of reionization scenario, bright galaxies should contribute by more than 50\%. In GN~42912-NE, the dust attenuation is slightly less, but still a significant sink, which contribution already adds to the what the neutral hydrogen absorbs. While the constraints from the photoionization models are slightly larger than for the doublet ratio method, all constraints are consistent within errors and suggest that the contribution of  GN~42912-NE is relatively low, at maximum 20\% in the conservative case (SMC law + taking the largest constraints), and likely lower than 10\% when considering the median of the different estimates obtained. 

% \begin{figure}
%     \centering
%     \includegraphics[width = \hsize]{figures/MGIIPhot_both.pdf}
%     \caption{Caption}
%     \label{fig:mgIIphot}
% \end{figure}

% R2796 = 0.079*lO32Corr_a**2  - 1.04*lO32Corr_a - 0.54 
% R2803 = 0.098*lO32Corr_a**2  - 1.02*lO32Corr_a - 0.84
 
% \subsection{Stack and upper limits}
% \label{sec:mgIInondeyt}

\section{A plausibly weak ionizing leakage in GN~42912-NE and GN~42912-SW} 
\label{sec:robustness}
In the previous section, we used \ion{Mg}{ii} to estimate \fesclyc\ for both GN~42912-NE and GN~42912-SW, considering instances where both galaxies are either dust free (NDC) or have dust extinction scaling with the measured $\beta$ values for both objects ($\beta$DC). Across these scenarios, we found a wide range of \fesclyc\ estimates for both objects. While this extensive range might initially appear to prevent definitive conclusions regarding the contribution of these galaxies during the EoR, this section highlights several key factors supporting the idea that the LyC escape from these two objects is likely weak, if not negligible.

In Section~\ref{sec:dust}, we discuss the NDC and $\beta$DC assumptions.  Section~\ref{sec:doublet_cav} examines our estimates in light of the caveats associated with \ion{Mg}{ii}-based indirect methods, as highlighted in \citet{katz2022}. Section~\ref{sec:consensus} summarizes the discussions from Section~\ref{sec:dust} and Section~\ref{sec:doublet_cav} to establish final reliable \fesclyc\ upper limits for GN~42912-NE and GN~42912-SW.

\subsection{The uncertainties on dust}
\label{sec:dust}
We first discuss the validity of the NDC scenario versus the $\beta$DC scenario in Section~\ref{sec:dustscenario}, and then comment on the impact of the choice of the dust extinction law on \fesclyc\ in Section~\ref{sec:dustlaw}.

\subsubsection{NDC and $\beta$DC scenarios}
\label{sec:dustscenario}
The impact of dust on nebular lines is usually corrected using the Balmer ratios. In Section~\ref{sec:emission}, we highlighted that GN~42912-NE has  H$\gamma$/H$\beta$ larger, although within 1$\sigma$, than the theoretical ratio under typical Case B assumptions, while GN~42912-SW has a H$\gamma$/H$\beta$ ratio consistent with an absence of dust attenuation. Both cases suggest a dust-free nebular environment, which contrasts with the observed $\beta$ slopes of $-1.5$ and $-1.9$. Given the significant uncertainties on the Balmer ratios and recent studies highlighting "unphysical" Balmer ratios in high-redshift objects \citep{topping2024metalpoor, cameron2024nebular}, we opted not to use these ratios for deriving the extinction-corrected emission line fluxes. Instead, we considered two scenarios: one assuming no dust (NDC) and another where $\beta$ is used as a proxy for nebular dust extinction.

Unsurprisingly, the NDC scenario yields the highest estimates of LyC escape fractions, as the final escape fraction at 912\AA\ relies solely on absorption from the neutral gas. However, the NDC scenario should be considered an extreme and likely unrealistic case. Indeed, an absence of dust attenuation for galaxies with $M_\star \sim 10^{8.5}M_\odot$ and 12+log(O/H)$\sim$8 seems highly unlikely since such galaxies should have had time to form metals and dust particles.

A dust-free scenario is also inconsistent with the $\beta$ slopes of GN~42912-NE and GN~42912-SW ($-1.92$ and $-1.51$, respectively). $\beta$ slopes are typically intertwined with both the age of populations emitting the intrinsic stellar continuum and dust extinction. However, \citet{chisholm2022_beta} demonstrated that for galaxies dominated by relatively young stellar populations, $\beta$ predominantly correlates with the amount of dust extinction in the galaxy. Using the remarkably tight $\beta$-to-$E(B-V)$ relation observed at $z\sim0.3$ \citet{chisholm2022_beta}, we show that the $\beta$ slopes of GN~42912-NE and GN~42912-SW should correspond to $E(B-V)$ values of approximately $\sim$0.134 ($\sim$0.043) and 0.204 (0.066), under the R16 (SMC) dust extinction law, suggesting relatively dusty environments.  

The $\beta$DC case is likely the most physically motivated scenario for these two objects. This scenario can also be considered conservative, as the nebular dust extinction is typically larger than the stellar dust extinction along the line of sight, traced by $\beta$ \citep[by a factor $\sim0.44$, ][]{calzetti2000}. Following Equation~\ref{eq:r2796}, adopting a larger nebular dust extinction would yield larger predicted intrinsic \ion{Mg}{ii} flux, hence leading to lower relative escape fractions and absolute escape fractions overall.

\subsubsection{On using different dust extinction curves}
\label{sec:dustlaw}

In this study, we also separately evaluated \fesclyc\ estimates based on the R16 dust extinction laws \citep{reddy2016dustlaw} and based on the SMC extinction law \citep{gordon2003}. Due to significant differences in their shapes, this approach offers valuable insights into how the choice of the dust extinction curve, which remains largely uncertain for galaxies at $z>6$, influences \fesclyc\ values. The variations in final \fesclyc\ estimates when considering either law are relatively minor (e.g. 0.063 $\pm$ 0.020 versus 0.085 $\pm$ 0.025 for GN~42912-NE) and falling within $1\sigma$ uncertainties. As the attenuation at 912\AA\ (A(912\AA)=($E(B-V)\times k$(912\AA)) is larger for the R16 law, the \fesclyc\ estimates derived using this curve are always smaller than when using the SMC law \citep[see also discussion in ][]{saldana2022, saldana2023}. Given the absence of constraints on the shape of the dust extinction curve for $z\sim7.5$ galaxies, hereafter, we use only the values derived using the SMC law since they are more conservative. Yet, opting for the R16-based estimates would not affect the discussion in the following sections.

\subsection{Caveats of indirectly predicting \fesclyc\ with \ion{Mg}{ii}}
\label{sec:doublet_cav}

Predicting LyC escape fractions indirectly is complex. LyC-diagnostic trends in the low-$z$ universe exhibit scatter over established empirical relationships \citep{flury2022_lycdiag}, and all indirect LyC diagnostics have caveats that must be carefully considered to understand the limitations of the estimates. \citet{katz2022_mgii} used radiation hydrodynamics simulations post-processed with radiative transfer to investigate a comprehensive list of caveats associated with \ion{Mg}{ii}-based approaches, which we discuss in this section. We address the limitations of the doublet ratio method in Section~\ref{sec:cavdoub} and explore the caveats of the photoionization models approach in Section~\ref{sec:cavphot}.

\subsubsection{Caveats of the \ion{Mg}{ii} doublet ratio method}
\label{sec:cavdoub}

The \ion{Mg}{ii} doublet ratio method suffers some caveats which are important to list here. First, as mentioned in Section~\ref{sec:doublet}, it relies on assumptions on the depletion fraction of Mg and of the Mg/H abundance ratio to perform the Mg$^+$ to H$^0$ conversion, but these ratios are not well known, especially at high-$z$. Using simulations, \citet{dubois2024_depletion} find that the Mg depletion factor can be significantly influenced by the gas density and the dust composition. Changes in the depletion factor or Mg/H abundance ratio will impact the inferred $N_{\rm \ion{H}{i}}$ and the escaping ionizing fraction derived from it.

Another caveat of the \ion{Mg}{ii} doublet ratio method is its reliance on the optically thin regime for accurate \fesclyc\ estimates. This approach is most reliable when the doublet ratio is close to 2 \citep{chisholm2020, chang2024_mgii}. However, as we move further into the optically thick regime, resonant radiative transfer effects become increasingly significant. These effects enhance the observed flux by scattering light from directions outside the line of sight into the line of sight and alter the shape of the doublet lines. This scattering process complicates the direct relationship between the doublet ratio value and the actual amount of neutral hydrogen gas present in the galaxy, making it more challenging to quantify the neutral gas content accurately.

When radiative transfer effects become significant, there is a risk of overpredicting the \fesclyc\ derived using the doublet ratio method. This effect is observed in the analysis of \citet{katz2022}, whose authors conducted tests on the reliability of this approach using galaxies from the \textsc{sphinx} cosmological simulation suite \citep{rosdahl2018}. For example, their Figure~17 emphasizes that predictions based on R(\ion{Mg}{ii}) can substantially overestimate the true escape fraction by $\sim$one order of magnitude. This overestimation is most significant for cases with low \fesclyc, which are cases where the neutral gas environment is overall optically thick.

In our analysis, we derived a doublet ratio of 1.28 for GN~42912-NE. Additionally, we find that the \ion{Mg}{ii} lines are slightly broader than other optical lines (150 km s$^{-1}$ for \ion{Mg}{ii} versus 100 km s$^{-1}$ for the [\ion{O}{ii}] lines), which is indicative of possible resonant scattering effects altering the line shapes. Both aspects support that we may not validate the optically-thin assumption for this galaxy (see also the discussion in Section~\ref{sec:proflowz}). Therefore, our $f^{\rm LyC}_{\rm esc,\ abs}$ estimates based on the doublet ratio may overpredict the true LyC escape in GN~42912-NE and therefore should be regarded as upper limits.

\subsubsection{Caveats of the photoionization-based method}
\label{sec:cavphot}

The photoionization-based method relies on \textsc{cloudy} models \citep{ferland2013, ferland2017} to deduce the intrinsic \ion{Mg}{ii} luminosity based on [\ion{O}{ii}] and [\ion{O}{iii}] optical lines. Initially proposed by \citet{henry2018} using idealized \textsc{cloudy} models, this approach was later refined and extended by \citet{xu2022_lya_mgII}. In our study, we utilized the \ion{Mg}{ii} to O$_{32}$ relation from the latter work to estimate the intrinsic flux for GN~42912-NE and GN~42912-SW.

However, using galaxies from the \textsc{sphinx} cosmological suite, \citet{katz2022_mgii} observed that, while the \ion{Mg}{ii} to O$_{32}$ relation proposed by \citet{henry2018} generally matches the simulated galaxies reasonably well, there is a certain scatter, with most \textsc{sphinx} galaxies tending to lie above the \citet{henry2018} relation. Consequently, predictions based on the latter relation could underestimate the \ion{Mg}{ii} intrinsic flux by $-0.2$ to $-1$ dex (see Figure~15 in \citealt{katz2022_mgii}), and therefore overestimate the relative escape fractions derived from these estimates. Given that the relations from \citet{xu2022_lya_mgII} are derived using a similar methodology, it is plausible that they may suffer from similar biases.

\citet{katz2022_mgii} proposed updated equations (their equations (3) and (4)) to derive a more accurate  \ion{Mg}{ii} intrinsic luminosity based on [\ion{O}{ii}] and [\ion{O}{iii}] lines. Using their equation in the $\beta$DC, we derived a relative escape fraction of ionizing photons of two times lower for GN~42912-NE. This aspect suggests that the photoionization-based \fesclyc\ estimates derived in Section~\ref{sec:photo} might overestimate both the true relative and absolute escape fractions for both objects.

\subsection{Reaching a consensus for realistic \fesclyc\ estimates in GN~42912-NE and GN~42912-SW. }
\label{sec:consensus}

So far, the discussion detailed in this section has highlighted three key points. First, the $\beta$DC scenario is likely more realistic than the NDC scenario to describe the environment of GN~42912-NE and GN~42912-SW. It is also a conservative way to estimate the nebular dust extinction since $\beta$ traces the stellar attenuation, typically lower than the nebular \citep{calzetti2000}. Assuming a larger nebular dust attenuation than in this work would only impact the dust-corrected flux values, which would be larger, hence yielding lower relative and absolute escape fractions than those reported in Table~\ref{sec:mgII}. 

%Note that the transformation from relative to absolute it would result in \fesclyc\ values lower .

Second, we found that the absolute \fesclyc\ estimates obtained using the SMC dust extinction law or R16 dust extinction law are consistent within $1\sigma$ in GN~42912-NE. Since the SMC law \citep{gordon2003} yields larger and more conservative \fesclyc\ estimates, it is, in the context of this work, a more prudent approach to setting estimates of the escape of ionizing photons in both galaxies. 

Finally, the doublet ratio from \citet{chisholm2020} and photoionization model methods based on \citet{henry2018} are consistent within 1$\sigma$. These two methods have been validated on LyC leaking galaxies observations at low-$z$, yet simulations at $z=6$ suggest that these methods may actually overpredict the relative escape fraction of LyC photons \citep{katz2022_mgii}.

Considering these three aspects collectively, we establish the final \fesclyc\ estimates for each galaxy as follows: for GN~42912-NE, we adopt the \fesclyc\ from the photoionization-based method with the SMC dust extinction law, resulting in a relative escape fraction of 31.0 per cent and an absolute LyC escape fraction of 8.5 per cent. Given the previously mentioned factors suggesting potential overestimation, we regard these values as upper limits. For GN~42912-SW, due to the lack of stringent constraints on the relative escape fraction, we apply a conservative upper limit on  $f_{\rm esc,\ rel}^{\rm LyC}$ of 100 per cent, leading to an upper limit on the absolute escape fraction of 14 per cent when using the SMC dust extinction curve. In the following section, we evaluate these \fesclyc\ constraints in the context of samples of leaking and non-leaking galaxies at $z\leq3$. We also examine these results alongside other indirect \fesclyc\ predictions obtained using alternative approaches from \citet{jaskot2024_lycmultiI}.

\section{Comparison with low-$z$ \ion{Mg}{ii} profiles and \fesclyc\ trends}
\label{sec:lyclit}

In the previous section, we established reliable upper limits on the escape fractions of ionizing photons for GN~42912-NE and GN~42912-SW at 8.5 per cent and 14 per cent, respectively. Here, we delve into how these estimates compare to observations of LyC leaking galaxies at low redshifts. Section~\ref{sec:proflowz} compares the \ion{Mg}{ii} profile of GN~42912-NE with those of a similar low-redshift leaking and non-leaking galaxy, Section~\ref{sec:trendlowz} places our \fesclyc\ estimates in context with trends observed in low-redshift samples, and Section~\ref{sec:compjaskot} explores the consistency of the \fesclyc\ estimates with alternative \fesclyc\ prediction models from \citet{mascia2023_fesc} and \citet{jaskot2024_lycmultiI}.

\subsection{Comparison to low-$z$ \ion{Mg}{ii} profiles}
\label{sec:proflowz}

\begin{figure*}
    \centering
    \includegraphics[width = 0.49\hsize]{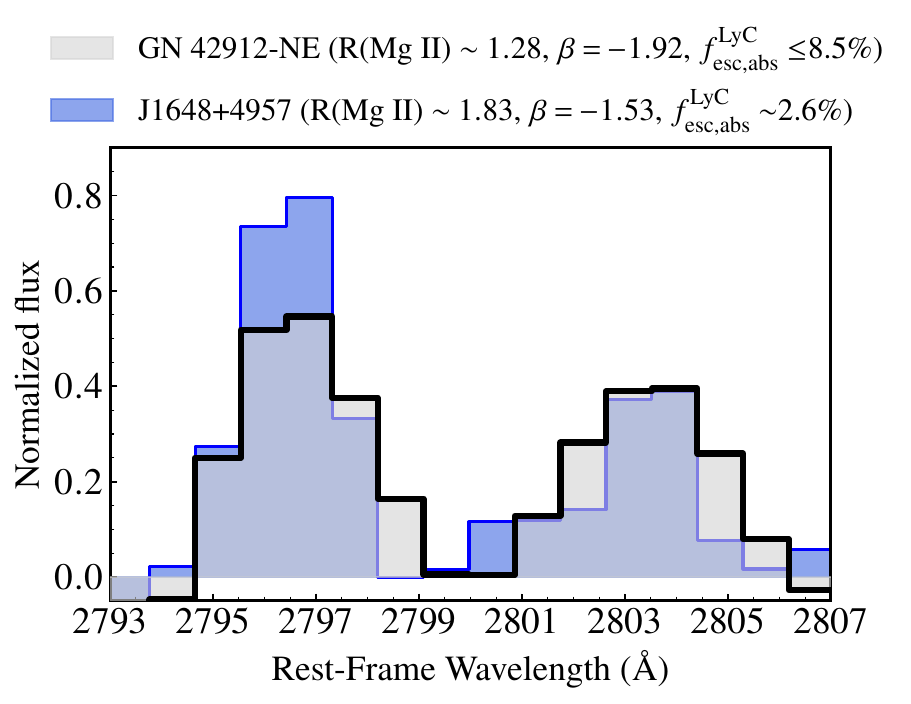}
    \includegraphics[width = 0.48\hsize]{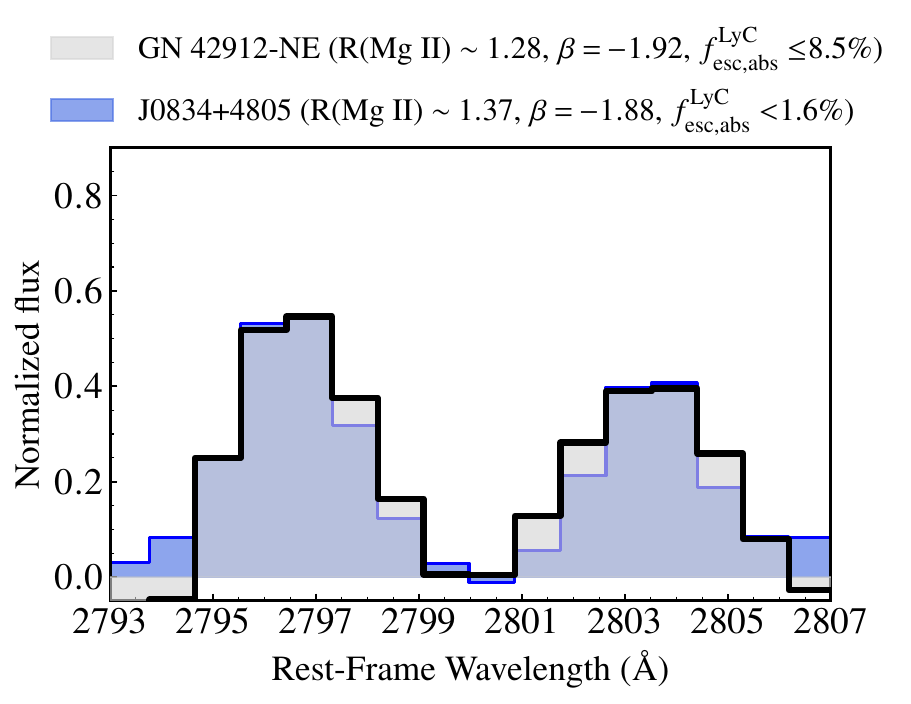}
    \caption{Comparison of the \ion{Mg}{ii} profile of GN42912-NE (in black) with that of two low-$z$ LyC leakers, J1648+4957 (left), and J0834+4805 (right) from the LzLCS sample \citep{flury2022_lzlcs}. The \ion{Mg}{ii} observations of these two low-$z$ galaxies (z $<$ 0.4) are from \citet{leclercq2024_lycMgII} and have been adjusted to the resolution of JWST-G235H. The \fesclyc\ values for the two LzLCS galaxies are determined using direct observations of the flux below 912\AA. J1648+4957 exhibits a \ion{Mg}{ii} doublet with a ratio of 1.83. J0832+4805 has a doublet ratio of 1.37, similar to that of GN42912-NE (1.28), and a similar $\beta$ slope ($-1.88$). The direct constraint on \fesclyc\ in this galaxy is $<1.6$ per cent, suggesting a negligible leakage. The comparison with these two low-$z$ galaxies with similar \ion{Mg}{ii} or $\beta$ properties emphasizes that GN~42912-NE likely has a weak or negligible LyC leakage. }
    \label{fig:mgiicomplow}
\end{figure*}

For over two decades, observations of LyC-leaking galaxies were exceedingly rare \citep{leitet2013, leitherer2016, borthakur2014}. A major breakthrough occurred with the detection of ten such galaxies in a series of Hubble Space Telescope observations, revealing significant \fesclyc\ values of up to 73 per cent \citep{izotov2016a, izotov2016b, izotov2018a, izotov2018b}. Building upon the lessons learned from these first detections, the LzLCS program (PI: Jaskot) \citet{flury2022_lzlcs} targetted a large number of LyC leaking candidates and nearly doubled the number of known LyC-leaking galaxies at $z<1$, providing a valuable sample for deriving indirect diagnostics relevant to high redshifts \citep{flury2022_lycdiag} and shedding light on the physical mechanisms behind the escape of ionizing photons \citep[e.g.][]{saldana2022, amorin2024, leclercq2024_lycMgII, jaskot2024_lycmultiI, wang2021_SII, bait2024_radiolzlcs}.

\citet{xu2023_lycmgII, xu2022_lya_mgII} and \citet{leclercq2024_lycMgII} presented the \ion{Mg}{ii} emission profiles of the LyC leaking and non-leaking galaxies from the LzLCS sample \citep{flury2022_lzlcs}. Here we compare the \ion{Mg}{ii} profile of GN~42912-NE with the \ion{Mg}{ii} profiles of two low-redshift galaxies from this sample. For a consistent comparison, we convolve the line profiles of the low-$z$ galaxies to the resolution of the NIRSpec-G235H instrument. 

In the left panel of Figure~\ref{fig:mgiicomplow}, we compare GN~42912-NE with J1648+4957, a $z\sim0.382$ galaxy with $M_{\rm UV} \approx -19.8$, log($M_\ast/M_\odot) \approx 8.5$, and 12+log(O/H) $\approx$ 8.25. Its escape fraction of LyC photons, determined directly using observations of the ionizing flux ($<912$\AA), is of 2.6 per cent. We chose this galaxy because its \ion{Mg}{ii} profile is typical of the profile of LyC leaking galaxies at low redshift.   R(\ion{Mg}{ii}) is 1.83, close to that optically thin limit of 2, and corresponds to a \ion{Mg}{ii} relative escape fraction of 49 per cent \citep[determined using the photoionization model approach;][]{xu2023_lycmgII}. The \ion{Mg}{ii} lines in GN~42912-NE are noticeably broader than those in J1648+4957, suggesting that resonant scattering significantly influences the line shape, while, in contrast, the \ion{Mg}{ii} lines in J1648+4957 likely validate the optically thin assumption. It is interesting to observe that despite an R(\ion{Mg}{ii}) close to the optically thin limit, the absolute \fesclyc\ of J1648+4957 is relatively low. This can be explained by the presence of a significant dust extinction of the ionizing photons as J1648+4957 is relatively red, with $\beta\sim-1.53$. This case particularly emphasizes that accurate \fesclyc\ constraints must account for both the neutral gas and the dust extinction along the line of sight.

The right panel of Figure~\ref{fig:mgiicomplow} compares the \ion{Mg}{ii} profile of  GN~42912-NE with that of J0834+4805 at $z\sim0.343$ with  $M_{\rm UV} \approx -19.9$,  log($M_\ast/M_\odot) \approx 9.1$ and 12+log(O/H) $\approx$ 8.17. Direct observations of the ionizing flux in J0834+4805 put constraints on \fesclyc$<1.6$ per cent \citep{flury2022_lzlcs}. The comparison with this galaxy is most interesting because J0834+4805 has striking similarities with GN~42912-NE, with a highly consistent \ion{Mg}{ii} profile (with R$\sim$1.37), a $\beta$ slope of $-1.88$, and an O$_{32}$ ratio of 4.3 as compared to 5.3 in GN~42912-NE. The similarities between both galaxies suggest that the \fesclyc\ upper limit established for GN~42912-NE is realistic, if not too conservative. Overall, the comparison to both low-$z$ galaxies further strengthens the idea that the \fesclyc\ in GN~42912-NE is likely weak if not absent.

% In that case,derived from the doublet ratio are close to 100\%, indicating that neutral gas plays a minor role in regulating the LyC escape. Rather, its \fesclyc\ is largely driven by dust extinction, which removes $\sim$97\% of the escaping radiation in J1648+4957 \citep[E(B-V)(R16) is 0.1636,][]{leclercq2024_lycMgII}.
% It is clear from Figure~\ref{fig:mgiicomplow} that the \ion{Mg}{ii} profile of GN~42912-NE differs notably from that of J1648+4957. As discussed earlier, the \ion{Mg}{ii} profile of GN~42912-NE suggests an overall relatively dense neutral gas environment, which acts as the major sink for ionizing photons. This observation further strengthens the notion that the \fesclyc\ in GN~42912-NE is likely weak, and possibly even absent.

Comparing higher-redshift objects to lower-redshift analogues comes with the assumption that both populations closely resemble each other. Although the first few years of {\sl JWST} observations have revealed a population of high-redshift objects with significantly different properties than those found in the local universe, (e.g. subdued dust contents, \citealt{cullen2024_dustfreepop}, or higher ionization, \citealt{bunker23_jades, tang2023jwstnirspecspectroscopyz79star}), multiple studies have pointed out the similarities in the spectral properties of high and low redshift star-forming galaxies \citep{schaerer2022_analogues, trump2023_analogues, rhoads_peas}. GN~42912-NE's \ion{Mg}{ii} features closely resemble those of galaxies in the nearby universe, reinforcing the conclusion that it is unlikely to be a dominant contributor to the ionizing photon budget of reionization. We note that, unlike the largely isolated LzLCS galaxies, GN~42912-NE may be interacting with GN~42912-SW. Such interactions can impact the escape fraction in opposing ways, either reducing it by triggering star formation in denser environments or enhancing it by creating low-density channels through dynamical processes. In this case, the observed \ion{Mg}{ii} features GN~42912-NE do not suggest a boosted ionizing escape.

\subsection{Comparison to lower redshift trends}
\label{sec:trendlowz}
\begin{figure*}
    \centering
    \includegraphics[width = \hsize]{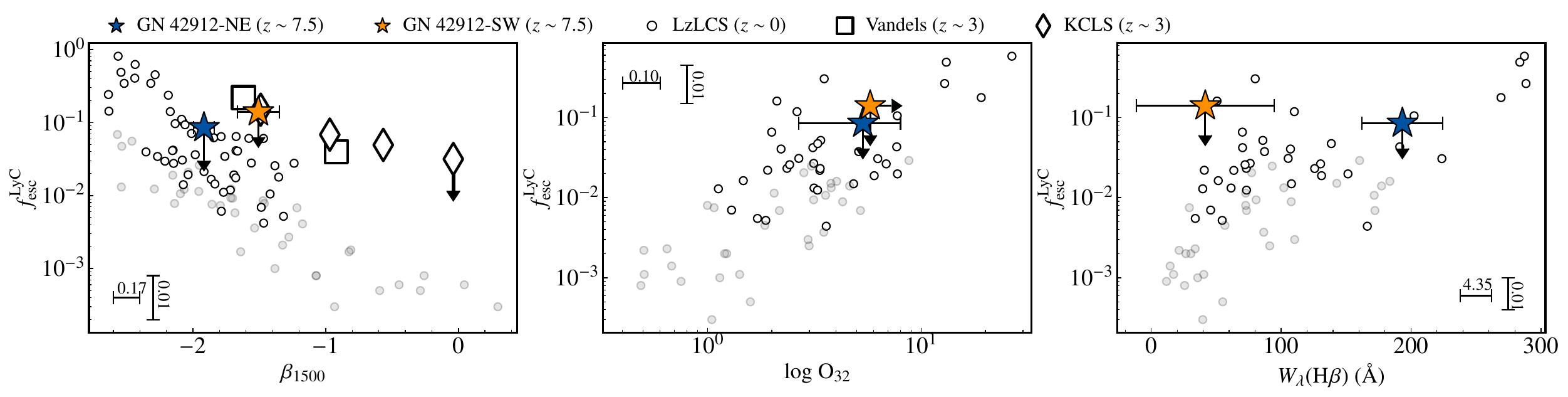}
    \caption{Comparison of the upper limits on \fesclyc\ for GN~42912-NE (blue) and GN~42912-SW (orange) with LyC trends with respect to stellar continuum slopes ($\beta$, left panel), the [\ion{O}{iii}] 5008\AA/[\ion{O}{ii}] 3727+3729\AA\ ratio (O$_{32}$, middle panel), and the H$\beta$ equivalent width ($W$(H$\beta$), right panel). The left panel includes the galaxies from the LzLCS \citep[$z\sim$0.3, with gray circles representing upper limits; ][]{flury2022_lzlcs}, VANDELS \citep[$z\sim$3; ][]{Begley2022_lyc}, and KLCS \citep[$z\sim$3;][]{steidel2018, pahl2021_lyc}, which encompass non-leakers, weak leakers, and strong leakers. Each panel includes typical uncertainties on the LzLCS data points. Overall, the upper limits on \fesclyc\ for GN~42912-NE and GN~42912-SW consistently align with the observed trends, suggesting that the GN-42912 system at $z = 7.5$ has similar LyC properties as lower redshift galaxies. }
    \label{fig:trendcomplow}
\end{figure*}

Here, we compare the upper limits on \fesclyc\ for GN~42912-NE and GN~42912-SW with LyC trends observed within the $z\leq3$ universe. We compare to publicly available properties of leaking and non-leaking galaxies from the LzLCS sample at $z\sim0.3$ \citep{flury2022_lzlcs}, the VANDELS sample at $z\sim 3$ \citep{McLure2018_vandels, Pentericci2018_vandels, Begley2022_lyc}, and the KCLS sample at $z\sim 3$ \citep{steidel2018, pahl2021_lyc}. In all three samples, the \fesclyc\ constraints are determined using spectral or photometric observations of the ionizing flux. All three samples have established significant correlations between \fesclyc\ and various properties such as Ly$\alpha$ profile properties, O$_{32}$, $\beta$, and the surface density of star formation $\Sigma_{\rm SFR}$. 

Here, we focus exclusively on the relationship of \fesclyc\ with three properties: the $\beta$ slopes at 1500\AA\ (measured from wavelengths between 1500 and 2000 \AA), the O$_{32}$ ratios, and $W$(H$\beta$). We note that the KCLS and VANDELS samples only appear in the \fesclyc-$\beta$ slopes comparison as we do not have the O$_{32}$ ratios and $W$(H$\beta$) for the same points. We also deliberately exclude \Lya\ from this comparison. This may seem surprising given that GN~42912 has a \Lya\ equivalent width constraint of $33 \pm 3$ Å from \citet{jung2020_lyatexas} and observational studies have shown strong correlations between LyC and \Lya\ properties. However, as discussed in Section~\ref{sec:overall}, the absence of \Lya\ detection in the PRISM observation of the same system from JADES indicates that \Lya\ emission may only be coming from one of the two components. The current observations do not provide enough information to determine which component it is emerging from. Therefore, we exclude \Lya\ from this section to avoid potential misinterpretation.

Figure~\ref{fig:trendcomplow} shows the three panels comparing the \fesclyc\ constraints of the GN~42912-SW and GN~42912-NE with $\beta$, O$_{32}$, and $W$(H$\beta$) trends. We observe that the \fesclyc\ upper limit derived for GN~42912-NE consistently aligns with the trends observed in the LzLCS sample. Regarding GN~42912-SW, its \fesclyc\ upper limit stands slightly above the $z\sim0.3$ $\beta$ and $W$(H$\beta$). This is not unexpected since we used the most conservative upper limit of 100 per cent on the relative escape fraction of LyC photons in this galaxy given the absence of \ion{Mg}{ii} and [\ion{O}{ii}] detection. Assuming a $f_{\rm esc,\ rel}^{\rm LyC}$ of 50 per cent instead would have yielded an absolute escape fraction to 7 per cent, aligning with the observed trends.  

The comparison between \fesclyc\ constraints and the $\beta$ slopes shown in the first panel of Figure~\ref{fig:trendcomplow}  presents interesting outcomes. This comparison includes data from the three samples of leaking and non-leaking galaxies at $z\sim0$ and $z\sim3$. As noted in previous studies \citep[e.g.][]{saldana2023}, the \fesclyc\ values of $z\sim3$ galaxies from the KCLS and VANDELS samples are significantly higher than those of low-$z$ galaxies from the LzLCS sample for similar $\beta$ slopes. This suggests that, for equivalent dust extinction properties, the $z\sim3$ galaxies allow more ionizing photons to escape compared to their low-redshift counterparts. This implies an evolution of the relative escape fraction of ionizing photons, potentially indicating a more neutral gas-depleted environment in these higher-redshift galaxies \citep[see also discussion in][]{saldana2023}. However, the \fesclyc\ upper limit of GN~42912-NE aligns consistently with the trends observed at $z\sim0$, while GN~42912-SW's upper limit is consistent with both the $z\sim0$ and $z\sim3$ trends. Although our data is limited and prevents a definite conclusion, these first two constraints may challenge the hypothesis of a simple linear redshift evolution of the relative escape fraction and, by extension, the neutral gas content. Further observations of LyC-leaking galaxies across different cosmic times are necessary to better understand the potential evolution of the relative escape fraction across redshifts.

In summary, Figure~\ref{fig:trendcomplow} shows that the \fesclyc\ upper limits for GN~42912-NE and GN~42912-SW align with trends observed in the $z<4$ universe, suggesting that low-redshift empirical relations may be applicable to high-redshift studies. However, the robustness of this comparison depends on how reliably high-redshift galaxies can be matched to their low-redshift counterparts. Uncertainties related to the shape of dust extinction laws, selection effects, and environmental factors, such as potential interactions within the GN~42912 system, limit our ability to draw definitive conclusions about the validity of these trends. Ultimately, only future observations and constraints on the \fesclyc\ of high-$z$ galaxies will bring clarity on this matter.

\subsection{Comparison to alternative \fesclyc\ predictions}
\label{sec:compjaskot}

Two recent studies proposed alternative approaches to indirectly estimating \fesclyc\ using galaxy properties that are accessible with {\sl JWST}. \citet{mascia2023_fesc} introduced a multivariate equation for determining \fesclyc\ based on $\beta$, the half-light radii ($r_e$), and the O$_{32}$ ratio. This equation was also adapted for cases where $W_\lambda$(H$\beta$) is measured instead of the O$_{32}$ ratio, since both quantities are tightly correlated \citep{mascia2024_feschighz}. More recently, \citet{jaskot2024_lycmultiI} utilized the Cox proportional hazards model \citep{cox1972} to generate multivariate models for \fesclyc. They found that models incorporating $\beta$ (or alternatively E(B-V)$_{\rm UV}$), O$_{32}$, and the surface density star formation rate ($\Sigma_{\rm SFR}$) successfully reproduced \fesclyc\ values in LyC leaking and non-leaking galaxies at $z\sim0.3$ and $z\sim3$. Both \citet{jaskot2024_lycmultiI} and \citet{mascia2023_fesc} developed their models using primarily the LzLCS sample and generally provide consistent \fesclyc\ estimates. However, \citet{jaskot2024_lycmultiI} noted that in the low \fesclyc\ regime (\fesclyc$<$0.05), the recipe from \citet{mascia2023_fesc} tends to overpredict \fesclyc. This mainly occurs because their linear multivariate analysis does not robustly account for non-detections and upper limits, while the Cox proportional hazards model does. 

% Second, \fesclyc\ predictions from \citet{mascia2023_fesc} underpredict \fesclyc\ for LyC leaking galaxies at  $z\sim3$, while the Cox models provide better matching constraints.

In this section, we compare the \ion{Mg}{ii}-based \fesclyc\ estimates with predictions derived using the models of \citet{jaskot2024_lycmultiI} and \citet{mascia2023_fesc}, utilizing the constraints on $\beta$, O$_{32}$, $\Sigma$SFR, and $r_e$ for GN~42912-NE and GN~42912-SW. For GN~42912-NE, the models from \citet{jaskot2024_lycmultiI} predict \fesclyc\ of 0.020$_{-0.016}^{+0.040}$, which is negligible and consistent with the upper limit derived using \ion{Mg}{ii}. Models from \citet{mascia2023_fesc} yield an expected leakage of 0.108$_{-0.0487}^{+0.288}$, which is higher but still consistent with the \ion{Mg}{ii}-based upper limit. The difference between the \fesclyc\ predictions from \citet{mascia2023_fesc} and \citet{jaskot2024_lycmultiI} is intriguing yet not unexpected since \citet{jaskot2024_lycmultiI} found similar discrepancies when $r_e$ is used instead of $\Sigma_{\rm SFR}$. Since \citet{jaskot2024_lycmultiI} predictions better reproduce the \fesclyc\ of $z\sim3$ LyC leaking galaxies, its prediction may more accurately represent the expected leakage in GN~42912-NE.

For GN~42912-SW, both approaches yield consistent estimates: \citet{jaskot2024_lycmultiI}'s model predicts \fesclyc\ of 0.005$_{-0.005}^{+0.023}$, while \citet{mascia2023_fesc}'s model predicts 0.026$_{-0.016}^{+0.041}$. These values are derived using the lower limit on O$_{32}$. Assuming an O$_{32}$ of 20 instead would increase the central value by approximately 3\% in both cases, maintaining minimal expected leakage. Overall, these estimates are consistent with the \ion{Mg}{ii}-based upper limit of 14\%. 

In general, the alternative approaches from \citet{jaskot2024_lycmultiI} and \citet{mascia2023_fesc} further support that both GN~42912-NE and GN~42912-SW exhibit weak or negligible ionizing photon leakage, aligning with the findings from the \ion{Mg}{ii}-based analysis.

\section{The expected contribution of GN~42912-NE and GN~42912-SW to reionization}
\label{sec:reio}

Using the \fesclyc\ constraints established in this work, we discuss here the expected impact of GN~42912-NE and GN~42912-SW on the neutral IGM (Section~\ref{sec:bubble}) and their \fesclyc\ constraints in the context of the \fesclyc-$M_{\rm UV}$ relations predicted by pre-JWST reionization models (Section~\ref{sec:model}). 

\subsection{Ionizing emissivity and bubble size}
\label{sec:bubble}

\begin{figure}
    \centering
    \includegraphics[width=\hsize]{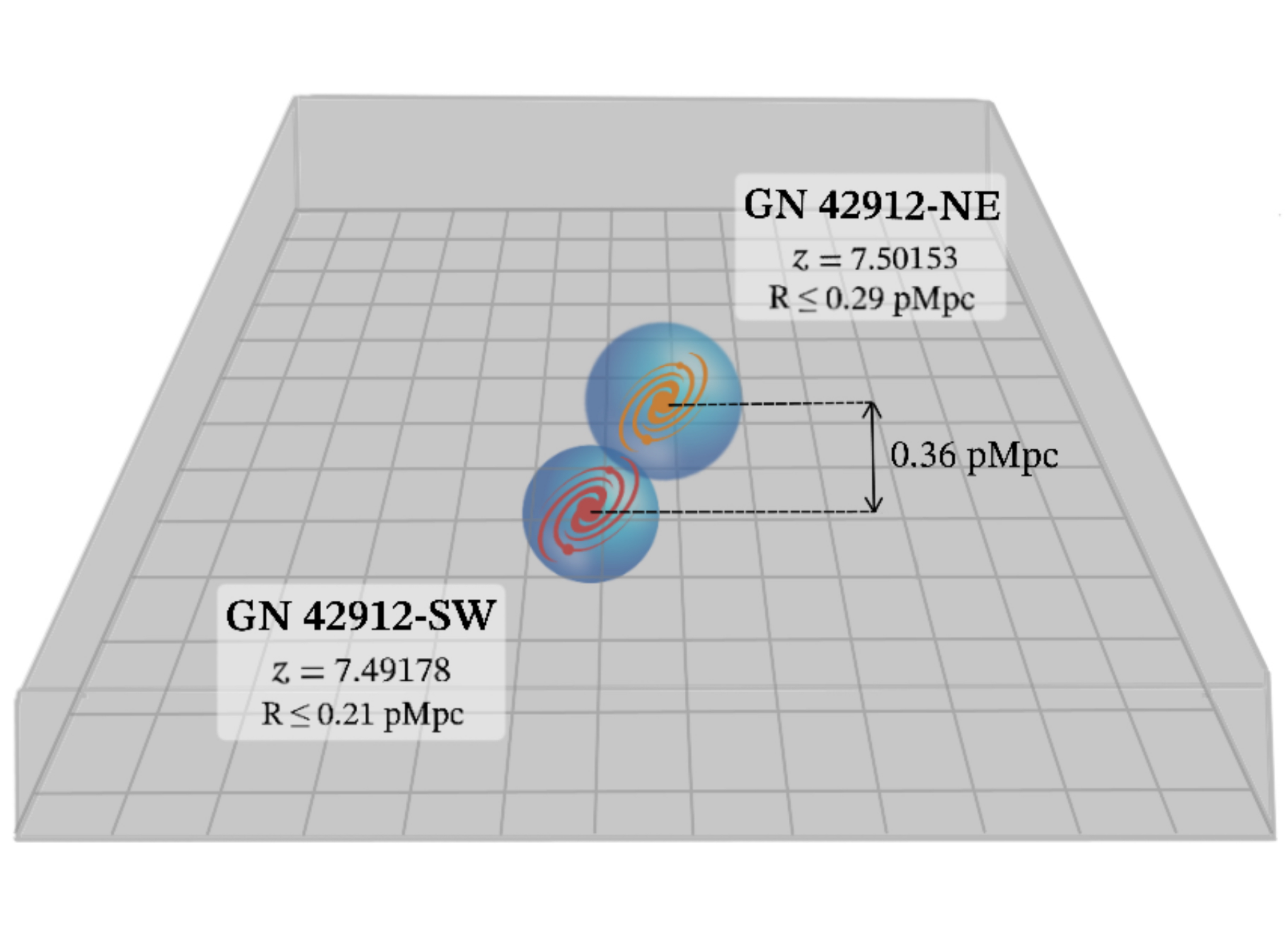}
    \caption{Simplified schematic of the GN~42912 system configuration, showing neutral hydrogen gas in gray and ionized hydrogen gas in blue. By using the H$\beta$ luminosity and \fesclyc\ constraints, we estimate the size of the ionized regions around both galaxies, finding that each creates a relatively small ionized bubble with a radius of approximately 0.17 pMpc. This suggests that both galaxies have a minimal impact on the surrounding neutral gas in the IGM.}
    \label{fig:bubble}
\end{figure}

The exact impact of galaxies on their surrounding medium depends on \fesclyc\ as well as the number of ionizing photons produced by the stars within the galaxy, $Q_{\rm H}^{\rm int}$. This parameter can be estimated using the H$\beta$ luminosity (L(H$\beta$)), as hydrogen recombination lines are relatively insensitive to metallicity and electron temperature \citep{charlot2001_nebular}. We follow the same approach described in \citet{Marques-Chaves2022_bluenugget}, using the equation that relates L(H$\beta$) to $Q_{\rm H}^{\rm int}$

\begin{equation}
    \text{L(H$\beta$)} = 2.1\times10^{12}\times Q_{\rm H}^{\rm int} \times (1-f_{\rm esc}^{\rm LyC}) \text{,}
\end{equation}

\noindent where  $2.1 \times 10^{12}$ is the nebular emission coefficient over the recombination rate in the Case B scenario \citep{Osterbrock2006}. Using the $\beta$DC scenario, we infer $Q_{\rm H}^{\rm int} = 6.43 \times 10^{53}$ s$^{-1}$ for GN~42912-NE and $1.95 \times 10^{53}$ s$^{-1}$ for GN~42912-SW. From $ Q_{\rm H}^{\rm int}$, we can directly infer an upper limit on $Q_{\rm H}^{\rm esc} $, the escaping rate of ionizing photons, which is simply $Q_{\rm H}^{\rm int} \times $ \fesclyc. We find $Q_{\rm H}^{\rm esc} \leq 5.46 \times 10^{53} $ s$^{-1}$ for GN~42912-NE and $ \leq 1.39 \times 10^{53} $ s$^{-1}$ for GN~42912-SW.

While the emission rate of ionizing photons provides valuable information, it is more insightful to translate this into an estimate of the size of the ionized regions that both objects can create during the EoR given their ionizing escape rate. This aspect has been explored in other studies \citep[e.g.][]{Larson_2022, endsley2021, Torralba2024_lyabubble}, particularly in the context of understanding the detection of \Lya\ at high redshift ($z>7$), where the IGM is expected to be predominantly optically thick. In such scenarios, \Lya\ detections may be attributed to the presence of sufficiently large ionized regions that allow the transmission of \Lya\ photons (typically $\geq$1 pMpc allow 50\% of the \Lya\ photons in the absence of kinematic effects, \citealt{mason_lya,endsley2022_lyaoffset, jung2024_ceerslya}). Following the methodology of \citet{Larson_2022, endsley2021, mason_lya}, which originates from \citet{cen2000_bubble}, we estimate the size of the ionized region around the two galaxies as

\begin{equation}
\text{R} =  \frac{3Q_{\rm H}^{\rm esc}t}{4\pi\langle n_{\rm H} \rangle} ,
\end{equation}
\noindent where $t$ is the duration of the current star formation phase and $\langle n_{\rm H} \rangle$  is the mean hydrogen density of the redshift of the source. Here we do not have robust constraints on the star-formation history so we adopt $t = 20$ Myrs, similarly as in \citet{Larson_2022}. 

Using the $Q_{\rm H}^{\rm esc}$ values inferred above, we estimate the sizes of the ionized regions around GN~42912-NE and GN~42912-SW to be approximately 0.29 pMpc and 0.21 pMpc, respectively (see Figure~\ref{fig:bubble}). Assuming a time span of 10 Myr, the sizes would be 0.23 pMpc for GN~42912-NE and 0.16 pMpc for GN~42912-SW. These regions are relatively small and indicate a minimal contribution to the ionization of the surrounding neutral IGM. However, the transverse separation between both objects is 3.1 cMpc, which is 0.36 pMpc at this redshift. This means that the ionized regions around both objects may overlap.

A typical ionized region size of 1 pMpc is required for significant Ly$\alpha$ photon transmission in the absence of kinematic effects \citep{mason_lya}. While the ionized bubbles around GN~42912-NE and GN~42912-SW are smaller than this threshold, \citet{jung2020_lyatexas} reported a significant Ly$\alpha$ detection with an equivalent width of 33.2\AA, indicating favorable conditions for Ly$\alpha$ photon propagation from at least one of these galaxies. We explore several scenarios that might explain the Ly$\alpha$ detection in the GN~42912 system.

One possibility is that fainter galaxies, which are currently undetected, could be contributing to the ionizing emissivity around both objects, potentially enlarging the actual size of the ionized region beyond what is accounted for by the ionizing photons escaping the two most luminous objects. However, \citet{endsley_jades} did not find evidence of a significant photometric overdensity around the GN~42912 system, which would support this hypothesis.

Another scenario to reconcile the significant Ly$\alpha$ emission detection is the overlap of their ionized bubbles. The combined maximum size of the regions around each object (0.29 pMpc and 0.21 pMpc) exceeds their physical separation at this redshift (0.36 pMpc). In this case, Ly$\alpha$ photons from GN~42912-NE could experience enhanced transmission due to the larger effective bubble size created by overlapping bubbles.

A third possibility is that the ionized bubbles were inflated by a previous burst of LyC leakage that has not yet recombined. Given the mass and metallicity of GN~42912-NE and GN~42912-SW, it is likely that these galaxies have undergone multiple star formation episodes. If an earlier phase of intense star formation led to significant LyC escape, the surrounding gas could have been ionized on a larger scale than expected from the current ionizing emissivity. In this scenario, the existing ionized regions may persist longer than the inferred current escape rates suggest, facilitating Ly$\alpha$ transmission even if the observed LyC leakage is weak.

Finally, kinematic effects may contribute to the Ly$\alpha$ detection. The transmission of Ly$\alpha$ photons depends on the velocity offset of the Ly$\alpha$ emission relative to the systemic velocity \citep{mason_lya, endsley2022_lyaoffset}. For a velocity offset of 400 km s$^{-1}$, an ionized region with a radius of 0.2 pMpc corresponds to a 40\% transmission rate, which drops to 20\% for a 750 km s$^{-1}$ offset. \citet{jung2020_lyatexas} determined that the peak of the Ly$\alpha$ emission is at $\lambda=1.034$ $\mu$m. This wavelength translates to a velocity offset of approximately 150 km s$^{-1}$ for GN~42912-NE and 500 km s$^{-1}$ for GN~42912-SW. If the Ly$\alpha$ originates from GN~42912-NE, the measured bubble size (not considering potential overlap) would only transmit 5\% of the escaping Ly$\alpha$ photons, necessitating a significant intrinsic Ly$\alpha$ emission to explain the detection. Conversely, if Ly$\alpha$ originates from GN~42912-SW, the 500 km s$^{-1}$ velocity offset results in about a 30\% transmission rate within an ionized region of 0.2 pMpc, making the Ly$\alpha$ detection plausible in this scenario.

Overall, further spatial Ly$\alpha$ mapping is required to pinpoint the origin and physical mechanisms behind the significant Ly$\alpha$ detection in the GN~42912 system. Importantly, the derived limits on the ionizing escape rates and ionized regions support the conclusion that both galaxies contribute minimally to the ionization of the surrounding neutral gas, emphasizing their weak role in reionizing the IGM hydrogen.

\subsection{the \fesclyc\ constraints of GN~42912 in the context of state-of-the-art reionization models  }
\label{sec:model}

 % In theory, understanding the drivers of reionization is relatively straightforward: we need to identify the astronomical objects that contribute the most to the ionizing photon budget and enable the ionization of neutral hydrogen by $z\sim6$ \citep{fan2006, dayal2018}. AGN were once possible candidates for reionization \citep{madau2015} but recent {\sl JWST} observations suggest they play a sub-dominant role \citep{matthee2024, dayal2024_agn_reio}, although this is still in debate \citep{madau2024_agn}. On the other hand, several pre- and post-{\sl JWST} studies have put star-forming galaxies, rather than AGN, as the prime candidates for driving the hydrogen reionization \citep[e.g.][]{robertson2015, finkelstein2019, Atek2024_epsion}. Two primary and simplistic models have emerged: in the first one, reionization is dominated by relatively faint objects (M$_{\rm UV}>-15$) with escape fractions between $\sim5$ and 10 per cent  \citep[e.g][]{finkelstein2019}, while in the other model, reionization is dominated by fewer but brighter galaxies (M$_{\rm UV}<-19$) exhibiting a higher average escape fraction of ionizing photons, \fesclyc\ $\sim 10$ to 20 per cent during the EoR \citep{naidu2020,naidu2022_lya50, matthee2022_lyc}. Both models yield predictions that are consistent with current constraints on the reionization history as well as results that are consistent with the Planck $\tau_{\rm CMB}$ estimates. 

Understanding reionization from the perspective of star-forming galaxies would have been straightforward in the perspective of the pre-{\sl JWST} era because many past theoretical studies assumed that star-forming galaxies produced stars at a nearly constant rate that produced a single, subdued $\xi_{\rm ion}$ value. In this context, one only needs to constrain the ionizing emissivity of both faint and bright galaxies and identify which sources dominated the ionizing photon budget. However, recent {\sl JWST} observations have significantly altered our understanding of reionization with the observations of early galaxies with remarkably high ionizing efficiencies \citep[${\rm log}\ \xi_{\rm ion}$ $\geq$ 25.5][]{Atek2024_epsion, simmonds2024, endsley2023, prieto-lyon2023, hsiao2024, pahl2024spectroscopicanalysisionizingphoton}, alongside highly star-forming galaxy populations at $z>9$ \citep{finkelstein2023_sfrreio, harikane2023_earlyjwst, eisenstein2023_jades}. Both aspects pose challenges to theoretical ionizing budgets, suggesting an extremely rapid reionization ending at $z>8.5$ \citep{munoz2024, Atek2024_epsion}. This rapid reionization timeline conflicts with the $\tau_{\rm CMB}$ measurements from Planck \citep{planck2021} and observations of damped \Lya\ wings in quasars at $z>6$ \citep[e.g][]{greig2019_lyadamping}.

\citet{munoz2024} explores potential solutions to this ionizing photon budget crisis, proposing that the average \fesclyc\ of galaxies could be lower (around 3 per cent) than previously anticipated from models based on lower redshift observations. In this scenario, the high ionizing efficiencies of star-forming galaxies would not contradict current constraints on the reionization history, because the actual amount of ionizing photons escaping from these galaxies is minimal. This scenario, already hinted at by the observations of galaxies with unexpectedly low escape fractions at $z<3$ \citep{Jung2024_lowescape}, is also supported by cosmological simulations. \citet{rosdahl2022_lyc} predict that galaxies with masses and metallicities similar to GN~42912-NE and GN~42912-SW contribute less than $\sim$15–20 per cent of the total ionizing budget, with typical \fesclyc\ values of only a few per cent, consistent with our findings. If high-redshift galaxies are embedded in denser neutral gas environments than previously assumed, their effective \fesclyc\ would be suppressed, limiting their overall contribution to reionization.

In this context, constraining the \fesclyc\ of high-$z$ objects is not only crucial for understanding the contributions of faint and bright galaxies to reionization but has now become essential to determine if these objects exhibit sufficiently low LyC leakage to reconcile theoretical models with current observations. In this section,  we compare our \fesclyc\ estimates for GN~42912-NE and GN~42912-SW against pre-{\sl JWST} reionization models and the ongoing ionizing photon budget crisis.

\begin{figure}
    \centering
    \includegraphics[width = \hsize]{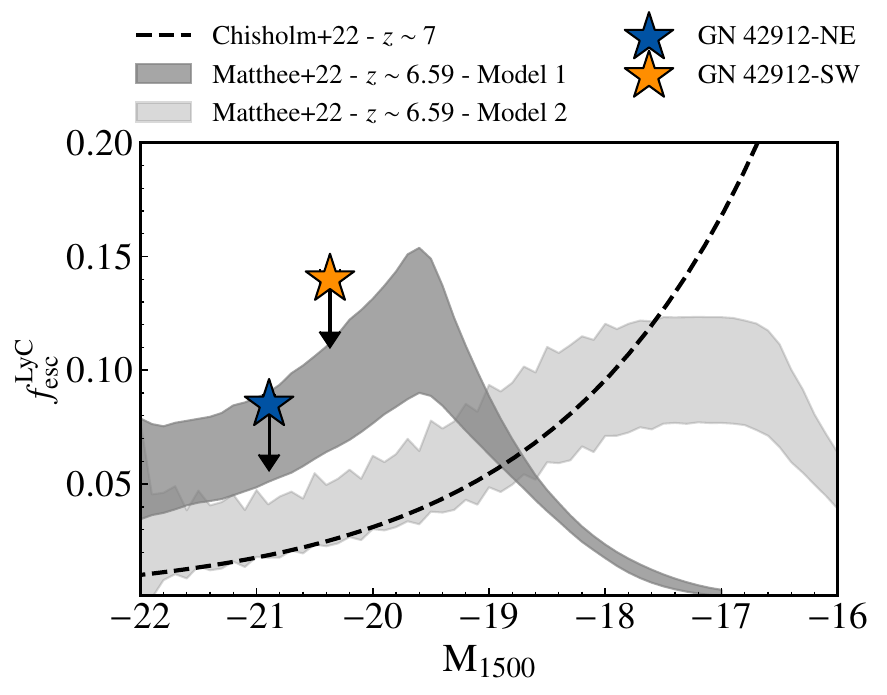}
    \caption{The evolution of \fesclyc\ around $z=7$ as a function of UV magnitude within the context of three reionization models and the $z\sim7.5$ \fesclyc\ constraints established in this work. The model proposed by \citet{chisholm2022_beta}, which relies on the $\beta$-to-\fesclyc\ relation identified in the LzLCS sample and the $\beta$–$M_{\rm UV}$ relation established by \citet{bouwens2014_uvcont}, argues for the dominance of relatively faint galaxies at high redshifts. The other two models are from \citet{matthee2022_lyc}, positing that reionization is dominated by a sub-population of luminous LAEs. Model 1 assumes half the LAEs have \fesclyc = 50 per cent, while Model 2 assumes half the LAEs have \fesclyc = 25 per cent. The \fesclyc\ constraints for GN42912-NE and GN42912-SW are not tight enough to definitively favor one model over the others. However, the relatively weak \fesclyc\ values for GN42912-NE and GN42912-SW suggest that these galaxies do not resemble the bright LAEs with \fesclyc = 25 per cent or 50 per cent that could explain reionization in the two \citet{matthee2022_lyc} models.}
    \label{fig:magnotude}
\end{figure}

In Figure~\ref{fig:magnotude}, we present three pre-{\sl JWST} models describing the evolution of the average \fesclyc\ as a function of UV magnitude at 1500\AA. The first model is by \citet{chisholm2022_beta}, based on observations of LyC-leaking galaxies from the LzLCS sample \citep{flury2022_lzlcs}. This model builds on the $\beta$ to \fesclyc\ relation established in the latter sample and the evolution of the $\beta$–M$_{\rm UV}$ colour–luminosity relationship established by \citet{bouwens2014_uvcont}. Since fainter galaxies are empirically found to be bluer, the $\beta$ to \fesclyc\ relation suggests that fainter galaxies have higher \fesclyc.

The other two models are from \citet{matthee2022_lyc}, based on observations of bright Ly$\alpha$ emitters (LAEs) at $z\sim2$ with fiducial LyC escape fractions of around 50 per cent \citep{naidu2022_lya50}. These models posit that a sub-population of LAEs dominates the reionization budget, naturally reproducing the evolution of emissivity at $z<6$. Here we plot two of these models, Model 1 assumes that half of the LAEs have \fesclyc\ = 50 per cent, and Model 2 assumes \fesclyc\ = 25 per cent. This translates to the faintest LyC-contributing galaxy having a limiting \Lya\ luminosity of 10$^{42.2}$ erg s$^{-1}$ in Model 1, and 10$^{41.2}$ erg s$^{-1}$ in Model 2.

Figure~\ref{fig:magnotude} illustrates how these different models behave as a function of $M_{\rm UV}$. \citet{chisholm2022_beta}'s faint-galaxy model shows a steady increase with \fesclyc\ $>$ 5 per cent for $M_{\rm UV} > -19$. Model 1 from \citet{matthee2022_lyc}, with half of the LAEs having \fesclyc\ = 50 per cent, peaks at $M_{\rm UV} \sim -20$ with an average \fesclyc\ around 12 per cent. Model 2, with \fesclyc\ = 25 per cent, peaks at $M_{\rm UV} \sim -17$ with an average \fesclyc\ around 8 per cent. Interestingly, Model 2 and \citet{chisholm2022_beta}'s model, despite their distinct assumptions, are consistent up to $M_{\rm UV} \sim -18$. The main difference lies in the \fesclyc-$M_{\rm UV}$ relation in \citet{matthee2022_lyc}'s models, where the reionization budget is dominated by a sub-population of LAEs, leading to a peaked relation that depends on the assumed \fesclyc\ for the LAE sub-population.

 We overlaid the \fesclyc\ of GN~42912-NE and GN~42912-SW on Figure~\ref{fig:magnotude}. As conservative upper limits, Figure~\ref{fig:magnotude} demonstrates that these constraints alone do not offer conclusive evidence for favoring any model over the others. Yet, both models from \citet{matthee2022_lyc} assume that 50 per cent of the LAEs contributing to reionization have \fesclyc\ $\sim$ 25 per cent (Model 2) and 50\% (Model 1). The upper limits of $<8.5$ per cent for GN~42912-NE and $<14$ per cent for GN~42912-SW suggest these two galaxies do not resemble the luminous LAEs that could explain reionization in such models.

While the contributions of GN~42912-NE and GN~42912-SW to the early universe's reionization are expected to be minimal, these isolated constraints are insufficient to significantly favor a faint or bright galaxy-dominated reionization model. Accurately determining the most relevant reionization model likely necessitates establishing \fesclyc\ for galaxies with faint UV magnitudes, specifically around $M_{\rm UV} \sim -17$ mag, where the average \fesclyc\ predicted by each model is sufficiently different to be distinguishable. Future {\sl JWST} programs focusing on the faint end of the galaxy population may be key to tipping the balance toward a certain reionization model.

Finally, the \fesclyc\ constraints of GN~42912-NE and GN~42912-SW also fall short of addressing whether reionization-era galaxies have lower \fesclyc\ than those predicted by pre-{\sl JWST} models, a factor that could potentially resolve the current ionizing budget crisis \citep{munoz2024}. Indeed, the upper limits derived do not definitively exclude the possibility of \fesclyc\ being greater than 3-5 per cent. If bright galaxies indeed emit an excessive number of ionizing photons, our findings do not conclusively argue that a reduced \fesclyc\ would align these observations with the reionization history of the universe. In general, while this study represents an important initial effort to constrain \fesclyc\ at high redshifts, it underscores the critical need for additional, and preferably more precise, constraints on \fesclyc\ in high-redshift galaxies to reconcile existing models with new observational data.

\section{Conclusions}
\label{sec:conc}

In this study, we analyzed the {\sl JWST} NIRSpec high-resolution G235H and G395H observations of GN~42912, a bright \Lya\ system at $z\sim7.5$ initially reported by \citet{finkelstein2013_lyadet}. These observations were part of the {\sl JWST} program 1871 (PI: Chisholm), with the primary objective of detecting the \ion{Mg}{ii} $\lambda\lambda$2796,2803 doublet to indirectly estimate the escape fraction of ionizing photons, building upon methodologies established and calibrated in the low-$z$ universe by \citet{henry2018} and \citet{chisholm2020}.

The {\sl JWST} observations unveiled that GN~42912 comprises two galaxies, denoted as GN~42912-NE and GN~42912-SW, separated by 358 km s$^{-1}$ in velocity and 0$\farcs$1 on the sky ($\sim 0.5$ kpc). GN~42912-NE has a stellar mass of $\sim 10^{8.4}M_\odot$, a UV magnitude of $-20.37$ mag, $\beta\sim-1.92$, a moderate O$_{32}$ ratio of 5.3, and a gas-phase metallicity estimated at $\sim 7.85$ using metallicity calibrations based on R$_{23}$ and O$_{32}$. Meanwhile, GN~42912-SW features a stellar mass of $\sim 10^{8.9}M_\odot$, a UV magnitude of $-20.89$ mag, $\beta\sim-1.51$, an O$_{32}$ ratio exceeding 5.5, and a gas-phase metallicity estimated around $\sim 8.09$.

Our analysis primarily focused on the \ion{Mg}{ii} $\lambda\lambda2796,2803$ observations of GN~42912-NE to constrain the escape of ionizing photons in this object. GN~42912-NE exhibits clear $>3\sigma$ detections of both lines, enabling us to employ two distinct approaches to estimate \fesclyc, the \ion{Mg}{ii} doublet approach relying on the ratio of the two lines \citep{chisholm2020}, and a comparison to the \ion{Mg}{ii} intrinsic flux determined using photoionization models \citep{henry2018}. We found that the \fesclyc\ estimates determined from both approaches are consistent within $1\sigma$. We calculated several estimates of the LyC escape fraction considering scenarios with and without dust and using two dust attenuation laws (the SMC law by \citealt{gordon2003} and the \citealt{reddy2016dustlaw} attenuation law) to explore the consequent variations on the indirect \fesclyc\ measurements. We find that the scenario without dust yields large \fesclyc\ estimates ($\geq50$ per cent), yet GN~42912-NE has a $\beta$ slope of $-1.92$, rendering this scenario an extreme scenario. \fesclyc\ constraints based on a more realistic scenario where $\beta$ is used as an empirical tracer of the dust extinction yields absolute escape fraction estimates between 6 and 14 per cent.

The absence of \ion{Mg}{ii} and [\ion{O}{ii}] detections in GN~42912-SW prevents us from using the \ion{Mg}{ii} based indirect approaches to establish constraints on the relative escape fraction of LyC photons in this galaxy. We assumed an upper limit of 100 per cent on the relative escape fraction (i.e., an absence of neutral gas), which yields absolute escape fractions less than 14 per cent (8.9 per cent) once accounting for the dust extinction at 912\AA\ and using the SMC (R16) dust extinction law.

We addressed the caveats of using \ion{Mg}{ii} to estimate \fesclyc\ (Section~\ref{sec:doublet_cav}). Mostly, these caveats, found by analyzing the relation between \ion{Mg}{ii} and \fesclyc\ in simulations, suggest that the \fesclyc\ estimates derived using the \citet{chisholm2020} or \citet{henry2018} methods may be overpredicting \fesclyc\ and should be regarded as strict upper limits. We concluded that \fesclyc\ is realistically less than 8.5 per cent in GN~42912-NE and less than 14 per cent in GN~42912-SW. These two estimates are consistent with alternative \fesclyc\ prediction models from \citet{mascia2023_fesc} and \citet{jaskot2024_lycmultiI} which also support a weak ionizing leakage in both objects (Section~\ref{sec:compjaskot}).

In Section \ref{sec:lyclit}, we compared these estimates to trends established from LyC leakers and non-leakers in the $z\leq3$ universe from the LzLCS \citep[z$\sim$0.3,][]{flury2022_lzlcs}, KCLS \citep[z$\sim$3,][]{steidel2018}, and VANDELS \citep[z$\sim$3,][]{McLure2018_vandels, Pentericci2018_vandels} samples. We showed that the \fesclyc\ values of GN~42912-NE and GN~42912-SW are consistent with the \fesclyc-$\beta$, \fesclyc-O$_{32}$, and \fesclyc-$W$(H$\beta$) trends seen at $z\sim0.3$. This suggests that local indirect tracers of LyC escape may be appropriate for observations of galaxies within the epoch of reionization. On the other hand, at fixed $\beta$, these $z\sim7.5$ \fesclyc\ constraints stand in line or lower than the $z\sim3$ constraints, suggesting that their neutral gas distributions may be more similar to $z\sim0.3$ leakers.

We derived the limit on the expected size of the ionized region around each object in Section~\ref{sec:bubble}. We found that the ionizing escape rate in GN~42912-NE corresponds to an ionized bubble with R$\leq0.29$ pMpc for GN~42912-NE and R$\leq0.21$ pMpc for GN~42912-SW, given the expected hydrogen density at $z=7.5$. These small inferred sizes further emphasize that both GN~42912-NE and GN~42912-SW likely have a minimal impact on the ionization of the hydrogen in the IGM at $z=7.5$.

Finally, Section~\ref{sec:reio} discusses these first $z\sim7.5$ \fesclyc\ constraints in the context of three reionization models, namely one model where relatively faint galaxies ($M_{\rm UV}>-19$) dominate reionization \citep{chisholm2022_beta}, and two models where a sub-population of luminous LAEs with \fesclyc=50 per cent or \fesclyc=25 per cent furnish the bulk of ionizing photons \citep{matthee2022_lyc}. We show that the current observations give limited constraints on the prevalence of one or the other models, but, given the weak expected leakage in these two luminous $z\sim7.5$ objects, these galaxies do not resemble the few potential bright objects that could explain reionization by having large \fesclyc.

% . Finally, we also consider our constraints in the context of the current JWST ionizing budget crisis \citep{munoz2024}. The \ion{Mg}{ii} profile of GN~42912-NE, hinting at large amount of hydrogen gas absorbing the escape radiation, may be consistent with new cosmological models which solve the tension by significantly decreasing the average \fesclyc\ of the contribution galaxies to $\sim3\%$, where this decrease being due to an overdensity of hydrogen within galaxies efficiently absorbing the LyC photons.

This paper presents the first \ion{Mg}{ii}-based constraints on the \fesclyc\ of galaxies during the EoR. While these constraints alone are insufficient to significantly determine the overall contribution of luminous galaxies to reionization, they represent an initial step by indicating a weak or possibly negligible leakage from two relatively bright galaxies at $z \sim 7.5$. Given the current context where {\sl JWST} has revealed galaxies with exceptionally high intrinsic production of ionizing photons, further \fesclyc\ constraints are crucial. These will help to assess whether the actual ionizing photon leakage from galaxies in the EoR is lower than previously anticipated by pre-{\sl JWST} models.

\section*{Acknowledgements}
This work is based on observations made with the NASA/ESA/CSA
James Webb Space Telescope. The data were obtained from the
Mikulski Archive for Space Telescopes at the Space Telescope Sci-
ence Institute, which is operated by the Association of Universities
for Research in Astronomy, Inc., under NASA contract NAS 5-03127
for JWST. These observations are associated with program \#01871.
Support for program \#01871 was provided by NASA through a grant
from the Space Telescope Science Institute, which is operated by the
Association of Universities for Research in Astronomy, Inc., under
NASA contract NAS 5-03127. SG is grateful for the support enabled by the Harlan J. Smith McDonald fellowship. Y.I. and N.G. acknowledge support from the Simons Foundation and the National Academy of Sciences of Ukraine (Project 0121U109612). ASL acknowledges support from Knut and Alice Wallenberg Foundation.

%%%%%%%%%%%%%%%%%%%%%%%%%%%%%%%%%%%%%%%%%%%%%%%%%%
\section*{Data Availability}
The data underlying this paper is available upon reasonable request.



\bibliographystyle{mnras}
\bibliography{bibliographie} % if your bibtex file is called example.bib

%\begin{thebibliography}{99}
%\end{thebibliography}

%%%%%%%%%%%%%%%%%%%%%%%%%%%%%%%%%%%%%%%%%%%%%%%%%%

%%%%%%%%%%%%%%%%% APPENDICES %%%%%%%%%%%%%%%%%%%%%

\appendix

\bsp	% typesetting comment
\label{lastpage}
\end{document}